\documentclass[fleqn,usenatbib,useAMS]{mnras}

\usepackage[british]{babel}
\usepackage{graphicx}	
\usepackage{amsmath}	
\usepackage{amssymb}	

\newcommand{\mpl}{M_{\rm Pl}} 
\newcommand{\dd}{\mathrm{d}}     
\newcommand{\Msun}{\,h^{-1}\,{\rm M}_{\sun}} 
\newcommand{\mpch}{\,h\,{\rm Mpc}^{-1}}
\newcommand{\mpcch}{\,h^{3}\,{\rm Mpc}^{-3}}
\newcommand{\hmpc}{\,h^{-1}\,{\rm Mpc}}

\newcommand{\hkpc}{\,h^{-1}\,{\rm kpc}}

\newcommand{\hgpcc}{\,h^{-3}\,{\rm Gpc}^3}
\newcommand{\lcdm}{\Lambda\mathrm{CDM}}
\newcommand{\CR}{\mathcal{R}}
\newcommand{\HP}{\mathcal{H}}

\usepackage[T1]{fontenc}
\usepackage{ae,aecompl}

\usepackage{txfonts}    

\hypersetup{pdfauthor={Arnalte-Mur, Hellwing and Norberg},
               pdftitle={Halo clustering in f(R) cosmologies},
               bookmarksnumbered=true}

\title[Halo clustering in $f(R)$ cosmologies]{Real- and redshift-space halo clustering in $f(R)$ cosmologies}

\author[Arnalte-Mur, Hellwing \& Norberg]
{
Pablo Arnalte-Mur, $^{1,3,4}$\thanks{E-mail: pablo.arnalte@uv.es}
Wojciech A. Hellwing $^{2,3,5}$
and  Peder Norberg $^{3,6}$
\\
$^1$Observatori Astron\`omic de la Universitat de Val\`encia, C/Catedr\`atic Jos\'e Beltr\'an, 2, 46980 Paterna,  Spain\\
$^2$Institute of Cosmology and Gravitation, University of Portsmouth, Portsmouth PO1 3FX, UK\\
$^3$Institute for Computational Cosmology, Department of Physics, Durham University, South Road, Durham DH1 3LE, UK\\
$^4$Departament d'Astronomia i Astrof\'isica, Universitat de Val\`encia, 46100 Burjassot, Spain\\
$^5$Janusz Gil Institute of Astronomy, University of Zielona G\'ora, ul. Szafrana 2, 65-516 Zielona G\'ora, Poland\\
$^6$Centre for Extragalactic Astronomy, Department of Physics, Durham University, South Road, Durham DH1 3LE, UK
}

\date{Accepted for publication in MNRAS - 19th January 2017}
\volume{accepted}

\pubyear{2017}

\begin{document}
\label{firstpage}
\pagerange{\pageref{firstpage}--\pageref{lastpage}}
\maketitle

\begin{abstract}
We present two-point correlation function statistics of the mass and the halos in the
chameleon $f(R)$ modified gravity scenario using a series of large volume N-body simulations.
Three distinct variations of $f(R)$ are considered (F4, F5 and F6) and compared to a fiducial
$\lcdm$ model in the redshift range $z \in [0,1]$. 
We find that the matter clustering is indistinguishable for all models except for F4, which shows a significantly steeper slope.
The ratio of the redshift- to real-space correlation function at scales $> 20 \hmpc$ 
agrees with the linear General Relativity (GR) Kaiser formula for the viable $f(R)$ models considered.
We consider three halo populations characterized by spatial abundances comparable to that 
of luminous red galaxies (LRGs) and galaxy clusters. The redshift-space halo correlation functions of F4 and F5 deviate 
significantly from $\lcdm$ at intermediate and high redshift, as the $f(R)$ halo bias is smaller or 
equal to that of the $\lcdm$ case.
Finally we introduce a new model independent clustering statistic to distinguish $f(R)$ from GR: the 
relative halo clustering ratio -- $\CR$. The sampling required to adequately reduce the scatter in
$\CR$ will be available with the advent of the next generation galaxy redshift surveys. This will
foster a prospective avenue to obtain largely model-independent cosmological constraints
on this class of modified gravity models.
\end{abstract}




\section{Introduction}
\label{sec:intro}

The hot relativistic big-bang $\Lambda$-cold dark matter ($\lcdm{}$) cosmology 
is a very successful standard model of cosmology.
It passes a tremendous amount of observational tests, from properties of the CMB
\citep[e.g.][]{WMAP9}, large-scale clustering of galaxies 
\citep[e.g.][]{bao_2df,bao_sdss,Zehavi2011,Alam2016a}, 
weak and strong lensing \citep[e.g.][]{Bartelmann2001,Schrabback2010,Suyu2013}
to properties of galaxy clusters, galaxies and their satellites 
in the nearby Universe 
\citep[e.g.][]{Mandelbaum2006, Allen2011,Wojtak2011,Guo2015,Umetsu2016}.
The minimum set of  parameters describing this simple scenario has been now established
to a remarkable precision \citep{Planck2015}. Despite its undeniable success, 
the standard $\lcdm{}$ model suffers from serious theoretical problems.
The model explains the observed late-time acceleration of the Universe \citep{acceleration1,acceleration2}
by attributing it to a very low positive value of Einstein's cosmological constant, $\Lambda$.
One of the main shortcomings of this approach comprise the fact that the only known
possible physical explanation of the non-zero $\Lambda$ is the zero-point
energy of vacuum quantum fluctuations. However, quantum theory predicts 
a natural value for $\Lambda$ that is many orders of magnitude larger than
the actual value that is compatible with observations 
\citep[for an excellent discussion see, e.g.][and references therein]{Carroll2001}.
The unavoidable conclusion is that one of the fundamental ingredients of the equations 
describing the evolution of the cosmological background 
is lacking a clear physical interpretation. 
In addition General Relativity (GR), as any working physical theory, itself needs to 
be continuously tested on all scales and regimes accessible through experiments and observations \citep{Will2014}.

The conceptual problems of $\lcdm{}$ have motivated a number of theoretical 
modifications to the standard model, which can produce
the observed late-time acceleration of the Universe by means of different
physical mechanisms. The rich literature on the subject can be
divided broadly into two distinct categories. 
In the first, it is postulated,
that the acceleration is produced by a dynamically evolving
background scalar field \citep[for a solid review of the subject see][]{Copeland2006}. 
These models, usually invoking a minimally coupled scalar field,
are collectively dubbed as dark energy. 
The second category consists of theories
where the accelerated expansion is a manifestation of the modifications
to the Einstein-Hilbert action integral.
I.e. they implement modifications to the theory of GR, an otherwise
fundamental building block of modern cosmology \citep{bbds2008,Clifton2012,Koyama2016}. 
The latter class of models are the so-called modified gravity (MOG) models.
Here the late-time acceleration is fueled by extra terms
appearing in the cosmic Lagrangian and which act as an `effective Lambda' term.
Thus, in this approach a mechanism that would set the usual cosmological constant
to exactly zero is needed. Since such a mechanism has not yet been discovered,
this approach should not be regarded as an attempt to construct a new fundamental theory of gravity,
but rather an effort to probe the rich phenomenology of infrared modifications to GR,
with non-trivial effects on cosmological scales.
MOG models, in principle, can be constructed
in many different ways. In the recent years, one of the broadly investigated
models, that fall into the MOG category, is the so-called $f(R)$ gravity theory.
In this case the accelerated expansion is produced by an extra term
replacing $\Lambda$ in the action integral. This term consists of a non-linear
function $f$ taking as an argument the curvature scalar $R$ \citep{nv2007,dt2010,sf2010}.
This class of models exhibit rich and interesting new physics. In addition to
producing late-time acceleration they admit for a non-negligible fifth force
acting on small and intermediate cosmological scales 
(i.e. much smaller than the horizon, $\ll cH_0^{-1}$). 
This non-trivial and intrinsically non-linear fifth force can manifest 
itself in deviations of the large and small-scale clustering of galaxies and matter 
from the standard GR picture. 
In other words: in $f(R)$ gravity one can
have an universe exhibiting GR, or $\lcdm{}$-like, expansion history but
admitting, at the same time, a different history of growth of structures \citep{ftbm2007,bbds2008,olh2008,Li2013}.

Any potentially successful MOG theory is required to not only predict a global
expansion history compatible with observations, but also needs to pass
stringent local tests of gravity. The latter come from observed orbital
dynamics in the Solar System \citep[e.g.][]{Chiba2007,hs2007,sol_test}, pulsar timing \citep{Brax2014}
and as of recently the physics of gravitational waves
emitted during black hole mergers \citep{Raveri2015,Abbott2016}. 
See also \citet{Berti2015} for a discussion of other astrophysical test of modified gravity.
In the most general class of $f(R)$ theories the fifth-force can freely propagate whenever 
there is a gradient of the $f(R)$ scalar field (also called the {\it scalaron}). 
Thus, if this model 
wants to stay compatible with the local gravity tests, it needs to implement 
a mechanism for suppressing the fifth force in high-density regions, like
our Solar System or neutron star binaries. In $f(R)$ theories
this is accomplished by a convenient choice of the $f(R)$ function
that give rise to the so-called {\it chameleon mechanism} \citep{kw2004,bbds2008}. 
The chameleon mechanism makes the scalaron very massive in
spatial regions of high local curvature (density), this leads
to an effective suppression of any fifth-force propagation.
Contrastingly, in regions with a low local density,
the field is light and admits the propagation of the scalar fifth-force.
The effectiveness of the chameleon suppression is moderated by
the local density field. This makes this mechanism 
to be {\it intrinsically} environment dependent and thus highly non-linear in its nature.
Consequently, in this scenario, one can have  
regions of low cosmic density (such as e.g. cosmic voids)
in which the fifth-force strongly affects the dynamics and clustering
of galaxies, as well as regions with higher density,
where the theory can effectively behave as the classical GR.
As the degree of non-linearity in both matter and scalar cosmic fields 
increases fast during cosmic evolution, it quickly renders predictions
of simple linear and weakly non-linear perturbation theory unreliable \citep[e.g.][]{2015pta..conf...58H}.
Because of this, the use of N-body computer simulations is essential 
for forecasting reliable and accurate predictions. However the same
very non-linear nature makes such simulations much more challenging 
and more expensive than standard GR simulations.
In the recent years there has been a significant progress in
the development of modified gravity N-body solvers 
\citep[e.g.][]{oyaizu2008,svh2009,zlk2011,lztk2012,mog_gadget,Llinares2008,Llinares2013,Llinares2014,MOG_comp,Bose2016a}.
As an outcome, modern codes are not only capable of running big volume and high-resolution simulations, 
but also have attained the accuracy needed for the precision cosmology era of the current 
and forthcoming galaxy surveys, such as \textit{Euclid} \citep{Laureijs2011}, 
the \textit{Dark Energy Spectroscopic Instrument} survey \citep[DESI,][]{Levi2013}, 
or the \textit{Javalambre-Physics of the Accelerated Universe Astrophysical Survey} \citep[J-PAS,][]{Benitez2014a}. 
Thanks to this, it is now possible to study the galaxy, halo and matter clustering
properties of $f(R)$ gravity models with sufficient resolution.

In general, we can expect that in $f(R)$ models the modifications to GR
will manifest themselves as a modified history of growth of structures,
and thus will also affect the galaxy clustering and dynamics.
It has been shown in the literature that indeed this class of models
exhibit higher amplitude of matter power spectrum at small and
intermediate scales (i.e. $\la 20\hmpc$) \citep{olh2008,fr_bispectrum,Li2013},
and even on larger scales for the case of higher-order
clustering amplitudes \citep{Hellwing2013}.
Dark matter clustering in redshift space is also characterized
by stronger {\it Finger-of-God} (FOG) effects at small scales \citep{Jackson1972},
which is accompanied by more pronounced Kaiser effect \citep{Kaiser1987,Hamilton1992} 
at larger scales \citep{Jennings2012}.
The stronger FOG, which leads to more effective small-scale power damping
in redshift space, is a manifestation of dynamics enhanced by the fifth force.
This effective enhancement was also shown to be predicted,
as a prominent MOG 'smoking gun' feature, for the galaxy/DM halo
velocity field \citep{Hellwing2014PhRvL}. 
Other studies have shown that $f(R)$ models can lead to different predictions for density profiles and size of cosmic voids \citep{halosvoids_fr,Cai2015}, modified stellar evolution \citep{Sakstein2015}, or several characteristics of galaxy clusters: number counts \citep{svh2009}, X-ray or lensing radial profiles \citep{Wilcox2016a} and measured gas fractions \citep{Li2016b}.

All the above mentioned effects of MOG in general should manifest themselves in observations 
as deviations from the GR-based predictions. 
However the highly non-linear character of the galaxy formation process makes it very difficult 
to foster observational predictions with respect to GR/MOG differences. 
Highly energetic processes, such as star formation feedback and Active Galactic Nuclei (AGN) 
feedback affect the matter distribution up to scales of $20\hmpc$ \cite[e.g.][]{vanDaalen2011,vanDaalen2014,Hellwing2016}.
It was shown that, when matter clustering is concerned, the baryonic feedback effects are 
degenerate with enhanced clustering predicted by pure collisionless simulations of $f(R)$ \citep{mog_gadget}. 
Therefore a good strategy aimed to find a clean $f(R)$ signature is to look at both 
larger-scales and at more massive haloes. 
Here one can expect that
the baryonic effects should be relatively weaker, giving hope of reducing the baryonic-MOG effects degeneracy.

These previous works have studied the expected changes in the growth of structures in $f(R)$ 
models by analysing the changes in different clustering properties of the DM density field. 
However, in order to be able to compare the models with observational data from galaxy surveys, 
one needs to obtain a prediction for the clustering of galaxies.
This involves studying possible differences in the biasing mechanism between $\lcdm$ and the $f(R)$ models.
In principle this would require the modelling of the galaxy formation process in the $f(R)$ theory.
A first step in this direction is to study the clustering properties of DM haloes, as the bias 
of galaxies is closely related to the bias of the haloes in which they reside.
Moreover, when we restrict the study to the most massive haloes and linear or quasi-linear scales, the clustering of DM haloes is a good proxy for the clustering of the corresponding central galaxies.
A complementary approach was followed by \citet{He2016a} who used the sub-halo abundance matching technique to study the clustering of galaxies in the $f(R)$ model at small non-linear scales ($r \leq 6 \hmpc$).

The aim of this work is therefore to characterize the clustering properties of DM haloes in a set of 
$f(R)$ models and compare them to the $\lcdm$ model.
We explain how the clustering of massive haloes is affected by $f(R)$ enhanced dynamics in both 
real and redshift space. 
We also conduct our study for a range of cosmic times, aiming to find the epoch of cosmic evolution 
at which the relative differences between the models are strongest. 
Our ultimate goal is to confront the theoretical predictions with observations from galaxy redshift surveys.
Hence, when selecting our samples and defining our clustering observables, we try to match what 
could be feasible when using real data.
Following this approach, we define a new statistic that can be easily measured from observations, 
and which can potentially help discriminate between GR and $f(R)$ cosmologies in the real Universe.

This paper is organized as follows. 
In section~\ref{sec:theory-simus} we give a brief description of both the physical set-up of 
the $f(R)$ model and of the numerical simulations used in this work.
The clustering statistics and the definition of the different halo samples that we use are 
described in section~\ref{sec:descr-halo-clust}.
Section~\ref{sec:results} concerns the results of our analysis, while in section~\ref{sec:observ-tests}
we discuss potential observational clustering tests using the new clustering ratio statistic.
Finally in section~\ref{sec:conc} we give our conclusions.

\section{The \lowercase{$f$}$(R)$ gravity theory and simulations}
\label{sec:theory-simus}

Here we briefly introduce the physical set-up and basic
properties of the $f(R)$ modified gravity model accompanied
by a description of the numerical structure formation simulations 
used in this work.

\subsection{The \lowercase{$f($}$R)$ gravity theory}
\label{ssec:fr-gravity-theory}

The $f(R)$ gravity \citep{cddett2005} is an extension of GR that has been extensively studied 
in the literature in the past few years. 
The main properties of the model are widely known,
hence we will focus here on only a very brief introduction of this theory, 
referring the reader for more details to the rich literature on the subject 
\citep[see e.g.][for detailed reviews]{sf2010, dt2010}.

The theory is obtained by substituting the Ricci scalar $R$ in 
the Einstein-Hilbert action with an algebraic function $f(R)$,
\begin{equation}
\label{eq:fr_action}
S = \int{\rm d}^4x\sqrt{-g}\left\{\frac{\mpl^2}{2}\left[R+f(R)\right]+\mathcal{L}_{\rm m}\right\} \, .
\end{equation}
Here $\mpl$ is the reduced Planck mass, $\mpl^{-2}=8\pi G$, $G$ is Newton's constant, $g$ the determinant 
of the metric $g_{\mu\nu}$ and $\mathcal{L}_{\rm m}$
the Lagrangian density for matter and radiation fields (including photons, neutrinos, baryons and cold dark matter). 
By designing the functional form of $f(R)$ one can fully specify
a $f(R)$ gravity model.

Varying the action, eq.~(\ref{eq:fr_action}), with respect to the metric field $g_{\mu\nu}$, one obtains 
the modified Einstein equation
\begin{eqnarray}\label{eq:fr_einstein}
G_{\mu\nu} + f_RR_{\mu\nu} -g_{\mu\nu}\left[\frac{1}{2}f(R)-\Box f_R\right]-\nabla_\mu\nabla_\nu f_R = 8\pi GT^m_{\mu\nu},
\end{eqnarray}
where $G_{\mu\nu}\equiv R_{\mu\nu}-\frac{1}{2}g_{\mu\nu}R$ is the Einstein tensor, $f_R\equiv \dd f/\dd R$, $\nabla_{\mu}$ 
is the covariant derivative
compatible with $g_{\mu\nu}$, $\Box\equiv\nabla^\alpha\nabla_\alpha$ and $T^m_{\mu\nu}$ is 
the energy momentum tensor of matter and radiation fields. 
Eq.~(\ref{eq:fr_einstein}) is a fourth-order differential equation, but can also be considered 
as the standard second-order equation of GR with a new 
dynamical degree of freedom, $f_R$, the equation of motion of which can be obtained by taking 
the trace of eq.~(\ref{eq:fr_einstein})
\begin{eqnarray}\label{eq:fr_eom}
\Box f_R = \frac{1}{3}\left(R-f_RR+2f(R)+8\pi G\rho_{\rm m}\right),
\end{eqnarray}
where $\rho_{\rm m}$ is the matter density. This new degree of freedom $f_R$ is 
the scalaron mentioned earlier.

Our analysis here is mainly concerned with large-scale structures,
which are  much smaller than the Hubble scale. 
Since the time variation of $f_R$ is very small in
the models to be considered below, we shall work in the quasi-static limit by neglecting 
the time derivatives of $f_R$. It has been shown that by adopting this approximation,
the resulting modelled dynamics of the scalar and matter fields deviates negligibly
from the true dynamics \citep{Bose2015}.
Under this limit, 
the $f_R$ equation of motion, eq.~(\ref{eq:fr_eom}), reduces to
\begin{eqnarray}\label{eq:fr_eqn_static}
\vec{\nabla}^2f_R &=& -\frac{1}{3}a^2\left[R-\bar{R} + 8\pi G\left(\rho_{\rm m}-\bar{\rho}_{\rm m}\right)\right],
\end{eqnarray}
where $\vec{\nabla}$ is the three dimensional gradient operator, and the overbar takes
the background ensemble average of a quantity. 

Similarly, the Poisson equation, which governs the behaviour of the gravitational potential $\Phi$, simplifies to
\begin{eqnarray}\label{eq:poisson_static}
\vec{\nabla}^2\Phi &=& \frac{16\pi G}{3}a^2\left(\rho_{\rm m}-\bar{\rho}_{\rm m}\right) + \frac{1}{6}a^2\left[R-\bar{R}\right],
\end{eqnarray}
by neglecting terms involving time derivatives of $\Phi$ and $f_R$, and using eq.~(\ref{eq:fr_eqn_static}) to eliminate $\vec{\nabla}^2f_R$.

The above considerations foster two ways in which the scalaron field can affect cosmology: 
(i) the background expansion of the Universe can be modified by the new terms in 
eq.~(\ref{eq:fr_einstein}) and (ii) the relationship between the gravitational 
potential $\Phi$ and the matter density field is modified, which can affect the matter 
clustering and growth of density perturbations. Clearly, when $|f_R|\ll1$, we 
have $R\approx-8\pi G\rho_{\rm m}$ (see eq.~(\ref{eq:fr_eqn_static})) and thus 
eq.~(\ref{eq:poisson_static}) reduces to the usual Poisson equation; when $|f_R|$ is large, 
we will have rather $|R-\bar{R}|\ll8\pi G|\rho_{\rm m}-\bar{\rho}_{\rm m}|$ and then eq.~(\ref{eq:poisson_static}) 
simplifies to the standard Poisson equation, but with $G$ rescaled by $4/3$. 
The value $1/3$ is the maximum intensification factor of gravity in $f(R)$ models,
independent of the specific functional form of $f(R)$. 
The choice of $f(R)$, however, 
is crucial because it determines the scalaron dynamics 
and therefore when and on what scales the enhancement factor changes from 1 to $4/3$. 
Scales much larger than the range of the modification 
to Newtonian gravity mediated by the scalaron field ({\it i.e.}, the Compton wavelength 
of $f_R$) are unaffected and gravity is not enhanced there,
while on small scales, depending on the environmental matter density, the $1/3$ enhancement 
may be fully realized.
This results in a scale-dependent 
modification of gravity and therefore a scale-dependent growth rate of structures
already at the linear theory level \citep{Koyama2009}.

\subsubsection{The chameleon mechanism}
 
\label{subsect:fr_cham}

The gravity and Newtonian dynamics passes stringent tests coming from 
the Solar System observations, and so any $4/3$ force enhancement factor
related to $f(R)$ needs to avoid high-density regions as our Solar System.
The theory achieve this by implementing the so-called chameleon screening mechanism.

The basic idea of the chameleon mechanism is the following: the modifications to Newtonian gravity 
can be considered as a fifth force mediated by the scalaron field $f_R$. Because the scalaron 
is massive, this extra force experiences a Yukawa-type potential. Hence the enhanced gravity 
is decaying exponentially as $\exp(-mr)$, in which $m$ is the scalaron mass, as the distance 
$r$ between two test masses increases. In high matter density environments, $m$ is very
heavy and the exponential decay causes a strong suppression of the force over distance. 
In reality, this is equivalent to setting $|f_R|\ll1$ in high density regions because 
$f_R$ is the potential of the fifth force, and this leads to the GR limit as we have discussed above.

Consequently, the functional form of $f(R)$ is crucial in determining whether the fifth 
force can be sufficiently suppressed in high density environments. In this work we consider 
the $f(R)$ Lagrangian proposed by \citet{hs2007}, for which
\begin{eqnarray}\label{eq:hs}
f(R) = -M^2\frac{c_1\left(-R/M^2\right)^n}{c_2\left(-R/M^2\right)^n+1},
\end{eqnarray}
where $M^2\equiv8\pi G\bar{\rho}_{\rm m0}/3=H_0^2\Omega_{\rm M}$, with $H$ being 
the Hubble expansion rate and $\Omega_{\rm M}$ the present-day fractional
density of matter. Throughout the paper
a subscript $0$ always denotes the present-day ($a=1$, $z=0$) 
value of a quantity. It was shown by \citet{hs2007} that $|f_{R0}|\la0.1$ is 
already sufficient to pass the Solar system constraints, but the exact constraint depends on 
the behaviour of $f_R$ in galaxies and pulsating stars as well \citep{Sakstein2013,Sakstein2015}.
At the background level the scalaron $f_R$ always sits close to the minimum of the effective
potential, therefore for the smooth scalar field we have \citep{Brax2012}:
\begin{equation}
\label{eqn:background_approx_fr}
 -\bar{R}\approx 8\pi G\bar{\rho}_m - 2\bar{f(R)} = 3 M^2\left(a^{-3} + \frac{2c_1}{3c_2}\right)\,
\end{equation}
The Hu-Sawicki model we consider is fixed by requesting that the background expansion history
matches that of $\lcdm$. Thus, we set
\begin{eqnarray}\label{eq:c1c2ratio}
\frac{c_1}{c_2} = 6\frac{\Omega_\Lambda}{\Omega_{\rm M}}
\end{eqnarray}
where $\Omega_{\rm M}$ and $\Omega_\Lambda$ are respectively the present-day fractional 
energy densities of the matter and dark energy. The simulation we use in this work
use WMAP3 cosmological background parameters \citep{WMAP3} (see Table \ref{tab:prop-simus}).
Using $\Omega_\Lambda=0.76$ and $\Omega_{\rm M}=0.24$ and eq.~(\ref{eqn:background_approx_fr})
gives $|\bar{R}|\approx41M^2\gg M^2$ at late times.
Using this approximation simplifies the expression of the scalaron to the following form
\begin{eqnarray}
f_R \approx -n\frac{c_1}{c_2^2}\left(\frac{M^2}{-R}\right)^{n+1}.
\end{eqnarray}
The above considerations show that once a $\lcdm$ background is fixed, our chosen $f(R)$
model is completely specified by the two free parameters: $n$ and $c_1/c_2^2$. 
Henceforth, the ratio $c_1/c_2^2$ is also fixed by the averaged background value of the scalaron, $f_{R0}$,
at $z=0$. This yields
\begin{eqnarray}
\frac{c_1}{c_2^2} = -\frac{1}{n}\left[3\left(1+4\frac{\Omega_\Lambda}{\Omega_{\rm M}}\right)\right]^{n+1}f_{R0}.
\end{eqnarray}
Thus the choice $f_{R0}$ and $n$ fully specifies our model.

The particular $f(R)$ set-up we consider here have very interesting cosmological properties. At small scales in regions where
the local density is high the enhanced gravity will be suppressed and the dynamics will be Newtonian. Hence we can expect
that orbital satellites and halo close interactions will be very similar as in GR. However in regions exhibiting 
low densities, such as e.g. cosmic voids, the modified dynamics should affect both halo and galaxy clustering and velocities.
We specifically consider three flavours of the Hu-Sawicki $f(R)$ model with fixed $n=1$, that differ in the present-day
mean (background) scalaron value $|f_{R0}|=10^{-4},10^{-5}$ and $10^{-6}$. We dub the models F4, F5 and F6
consequently. These three models cover the portion of the $f(R)$ parameter space that produce interesting
cosmological effects and is still compatible with extragalactic observations.
While F5 and F6 are so far in a broad agreement with the cosmological observations, F4 however is already
in a strong tension with observations of cluster number counts \citep{svh2009,Ferraro2011,Lombriser2012,Cataneo2015}
or weak lensing \citep{Harnois-Deraps2015,Liu2016}.
Thus we shall use F4 results just as an extreme example
of effects induced by only weakly-screened fifth force. 

\subsection{Cosmological \lowercase{$f($}$R)$ simulations used in this work}
\label{ssec:simus}

In this work we use the $f(R)$ simulations introduced in \citet{Li2013}.
Most of the previous work however focused on DM density fields only. Here we are
very much interested in clustering properties of DM haloes (and ultimately galaxies).	
For that reason we have applied \textsc{Rockstar}, a phase-space Friends-of-Friends (FOF) halo finder \citep{Behroozi2013}.
We kept all the haloes that contained at least 100 DM particles, hence this sets
our minimal halo mass limit to $M_{\rm min}=2.09\times 10^{13}\Msun$. 
Further on we recompute the FOF halo mass using a proper virial mass
definition. For the virial mass we use $M_{200}$, i.e. the mass 
contained in a sphere of radius $r_{200}$ centred on a halo, 
such that the average overdensity inside the sphere is $200$ 
times the critical closure density, $\rho_c\equiv{3H^2/8\pi G}$.
Our adopted mass-cut left us with
$\sim 10^{6}$ haloes at $z=0$ for each initial condition realization. Thus 
the upper-limit on our spatial number density of objects is $\bar{n}=3\times 10^{-4}\mpcch$
at $z=0$ and correspondingly smaller at higher redshifts. 
Our simulations use a computational domain of $1500\hmpc$ size. 
Following the analyses by other authors of the importance of both finite-volume effects \citep[e.g.][]{Colombi1994}
and sparse-sampling \citep[e.g.][]{cosmic_error} we adopt conservative limits on the minimal
and maximal scales that we trust. For a minimum scale we adopt a limit
of $3\times 2\pi k_{\rm Nyq}^{-1}\simeq10\hmpc$, where the Nyquist frequency for the simulations is
$k_{\rm Nyq}=2.14\mpch$. 
We take as the maximum scale to study $1/10\times L_{\rm box}\simeq 150\hmpc$, as we expect larger scales to be affected by the finite volume effects.
Finally we will focus our analysis on 4 snapshots taken
consecutively at $z=0, 0.25, 0.66$ and $1.0$. Previous studies \citep{Hellwing2013} have shown
that for those times the differences between GR and $f(R)$ clustering
are expected to be the largest. We list other details of the simulations used here
in Table~\ref{tab:prop-simus}.

\begin{table}
  \caption{Main properties of the simulations used in this work, for more details please see \citet{Li2013, Hellwing2013}.}
  \label{tab:prop-simus}
  \centering
  \begin{tabular}{ll}
    \hline
    Models      &  $\lcdm$, F6, F5, F4 \\
    Number of realizations & 6 \\
    Box size             & $L_{\rm box} = 1500 \hmpc$ \\
    Number of particles  & $N_{\rm p} = 1024^3$    \\
    Particle mass        & $m_{\rm p} \simeq 2 \times 10^{11} \Msun$ \\
    Nyquist frequency    & $k_{\rm Nyq} = 2.14 \mpch$ \\
    Force resolution     & $\epsilon = 22.9 \hkpc$ \\ 
\hline
    Cosmological parameters:  & \\
    total matter density & $\Omega_{\rm M} = 0.24$ \\  
    dark energy density  &  $\Omega_{\Lambda}\; = 0.76$ \\
    baryonic matter density & $\Omega_{\rm b} \;= 0.04181$\\
    dimensionless Hubble parameter & $h\;\;\;\; = 0.73$ \\
    tilt factor of the initial power spectrum & $n_{\rm s}\;\; = 0.958$ \\
    power spectrum normalization & $\sigma_8\;\; = 0.77$ \\ 
    BAO peak scale (linear theory) &  $r_{\rm BAO}\simeq 113\hmpc$\\
\hline   
  \end{tabular}
\end{table}

\section{Analysis of halo clustering in $N$-body simulations}
\label{sec:descr-halo-clust}

The astronomical observations that provide the data characterizing the clustering
of matter at large scales contain information only about the luminous stellar matter distribution
in our Universe. Contemporary galaxy redshift catalogues contain positions of millions of galaxies,
observed over large parts of the sky and over vast distances (redshifts). Ideally one would like then
to study the clustering of galaxies in various competing cosmological models. This requires however
introduction of another component into a theory under investigation, namely the galaxy formation model.

Various techniques exist that allow for galaxy formation modelling, the semi-analytic models (SAMs) 
\citep[for a review see][]{Baugh2006},
hydrodynamical simulations \citep{Vogelsberger2014,Schaye2015} and abundance matching 
\citep{Kravtsov2004,Moser2010}, to just name a few. 
However all the existing techniques were developed and tested self-consistently only for the $\lcdm$ model. Application and extrapolation of such modelling
to MOG models is neither straightforward nor simple \citep{Fontanot2013}. In addition the existing $\lcdm$ galaxy formation models
are still subject of intensive scrutiny \citep{Contreras2013}, as our understanding of the importance and interconnection
of all the complicated baryonic feedback processes is far from being full and complete 
\citep[see e.g.][]{Schaye2010,McCarthy2010,Fabjan2010,McCarthy2011,Puchwein2013}.
In addition, the strength and the environmental dependence of the additional fifth-force of the $f(R)$ model 
impact the galaxy clustering in a way that is degenerated with strong baryonic feedback invoked
by AGNs and galaxy winds \citep{mog_gadget}.

Taking into account all the difficulties mentioned above and the challenges connected with galaxy formation, 
we decide to follow a simpler approach. 
We use DM haloes and their clustering properties as proxies for galaxy clustering. 
Haloes are well defined objects (both in $\lcdm$ and $f(R)$), and as such can be straightforwardly identified 
and extracted from N-body simulations \citep{Knebe2011}.

We expect that in $f(R)$ gravity the galaxy formation mechanism and processes involved can, in principle, take 
largely different character than in $\lcdm$. 
However, if we restrict the analysis to a sample of very luminous galaxies, the situation is simpler. 
In this case, the fraction of satellite galaxies is very small \citep[e.g.][]{Zheng2009a}, so we can assume that 
most of the galaxies are located at the centres of massive DM haloes and the galaxy clustering properties will 
follow closely those of the host haloes.
This is certainly a valid approximation if we constrain ourselves to sufficiently large scales (i.e. the two-halo 
term limit, $\geq10\hmpc$). 
Moreover, for this type of galaxy samples, it is possible to remove the effect of the satellite galaxies from 
clustering measurements \citep*{Reid2009a}.
Going further to the high-mass end of the mass function, DM haloes correspond to galaxy groups or clusters, 
which can be identified from galaxy surveys \citep[see, e.g.,][]{Koester2007a,Robotham2011a,Ascaso2015a}.
In this regime, the clustering of DM haloes in different models can be directly compared to that of observed groups.

To characterize the clustering of matter and haloes (in position and redshift space) at different scales and epochs we use a basic
2-point statistic: the two-point correlation function, $\xi(r)$. This is defined as \citep{Peebles1980}
the excess probability (with respect to a Poisson process) of finding two haloes contained in
two volume elements d$V_1$ and d$V_2$ at a distance $r$:
\begin{equation}
\label{eqn:2pcf-definition}
\dd P_{12}(r) \equiv \bar{n}^2[1+\xi(r)]\dd V_1\dd V_2\,,
\end{equation}
where $\bar{n}$ is the mean halo (galaxy) number density.

In general, the halo 2-point correlation function will depend on the selected halo population ($\HP$), the redshift ($z$),
and the cosmological model ($\mathcal{M}$) considered, which we denote as $\xi(r|z, \HP, \mathcal{M})$. 
Because a density perturbation in an expanding universe needs to pass a certain threshold value $\delta_c$
\footnote{The critical density threshold for collapse takes different values in various cosmologies.
For $\lcdm$ $\delta_c\simeq1.673$ \citep{Peebles1980, Weinberg2003a}. For $f(R)$ this number is no longer universal as the fifth-force has an environmental and scale dependence \citep{Li2012c}. 
\citet{sloh2009} have shown for example that, when the chameleon effect is ignored, the value for F4 is $\delta_c\simeq1.692$.}
in order to be able to collapse and form a gravitationally bound structure (i.e. a halo),
the haloes are biased tracers of the underlying smooth matter density field \citep[see e.g.][]{FG1993}.
We parametrize this through a simple linear relation:
\begin{equation}
  \label{eq:linear_bias}
  \xi(r|z, \HP, \mathcal{M}) = b^2(r|z, \HP, \mathcal{M}) \xi_{\rm m}(r|z, \mathcal{M}) \, ,
\end{equation}
where $b(r|z, \HP, \mathcal{M})$ is the linear bias parameter and $\xi_{\rm m}(r)$ is the correlation 
function of the matter density field.

Generally, we can expect that the main differences between $\lcdm$ and $f(R)$ halo clustering,
will arise due to: (i) different amplitudes of the matter correlation function, $\xi_m$, at the same scale $r$,
and (ii) deviation in the bias parameter, which will be driven by both the departure in the 
halo mass - bias relation and by the differences in the selection of a particular halo population.
We will study the clustering of haloes in redshift space, as this corresponds to what would be available from observations. 
Therefore, further differences can originate from changes in the effects of redshift-space distortions in different gravity models.

We present in section~\ref{ssec:estimation} the method we use to measure the halo correlation function in the simulations. In section~\ref{ssec:select-halo-popul} we show the halo mass functions obtained in the simulations and we also explain the approach used to select the different halo populations we analyse in section~\ref{sec:results}.

\subsection{Estimation of the correlation function in the simulations}
\label{ssec:estimation}

In this work, we estimate the correlation function for different tracers (haloes or DM particles) extracted from N-body simulations. 
This means that the selection function in all cases is complete, isotropic and homogeneous.
Moreover, as the volume is a box with periodic boundary conditions, we do not need to correct for any edge effects.
Therefore, we obtain the correlation function in each case using the simple estimator:
\begin{equation}
  \label{eq:2p_estimator}
  \hat{\xi}(r) = \frac{DD(r)}{N \bar{n} v(r)} -1 \, ,
\end{equation}
where $DD(r)$ is the number of pairs of tracers with separation in the range $[r, r+\Delta r]$, $N$ is the total 
number of tracers in the sample, $\bar{n}$ is their number density, and $v(r)$ is the volume of a spherical shell 
of radius $r$ and width $\Delta r$,
\begin{equation}
  \label{eq:vshell}
  v(r) = \frac{4 \pi}{3}\left[ \left(r+\Delta r\right)^3 - r^3 \right] \, .
\end{equation}
We use in all cases bins in separation of width $\Delta r = 8 \hmpc$.
This simple estimator is much faster than the estimators usually used for real data, such as that from \citet{Landy1993a}, 
as in this case we do not need to use an auxiliary random sample to correct for the selection function or edge effects. 
We checked that we obtained identical results when using the \citet{Landy1993a} estimator for our calculations.

We compute, for each tracer, the correlation function separately in each of our six realizations and take as our value 
for the correlation function of this tracer the mean of these six estimations. To estimate the corresponding error
we use the standard error on the mean over the ensemble of six realizations.
This is a conservative error estimation \citep{cosmic_error} that takes into account the contributions of both the 
cosmic variance and the shot noise.
Although cosmic variance is the main source of uncertainty for the dark matter correlation function, shot noise is 
also important when we consider samples of massive haloes, with low number density.
As we are combining here our six realizations, the statistical error we obtain would correspond to that achievable 
by an ideal survey covering a volume of $V = 6 \times \left(1500 \hmpc\right)^3 \simeq 20 \hgpcc$.
When we compute the dark matter correlation function (in section~\ref{ssec:matter-field}) we use a random subsample containing $\simeq N_p/1000$ DM particles in each realization.
This subsample is obtained by randomly selecting particles from the ID list, so that all the population properties are sampled uniformly.
This avoids the need for a prohibitive computation time, while not affecting the results, as the errors in 
the resulting sample are still dominated by cosmic variance, and not by shot noise.
For comparison, the resulting number density of DM particles used in our calculations is still $\sim10$ times 
larger than that of the densest halo sample used (see below).

\subsection{Halo mass function and selection of halo populations}
\label{ssec:select-halo-popul}

\begin{figure*}
  \centering
  \includegraphics[width=0.98\textwidth]{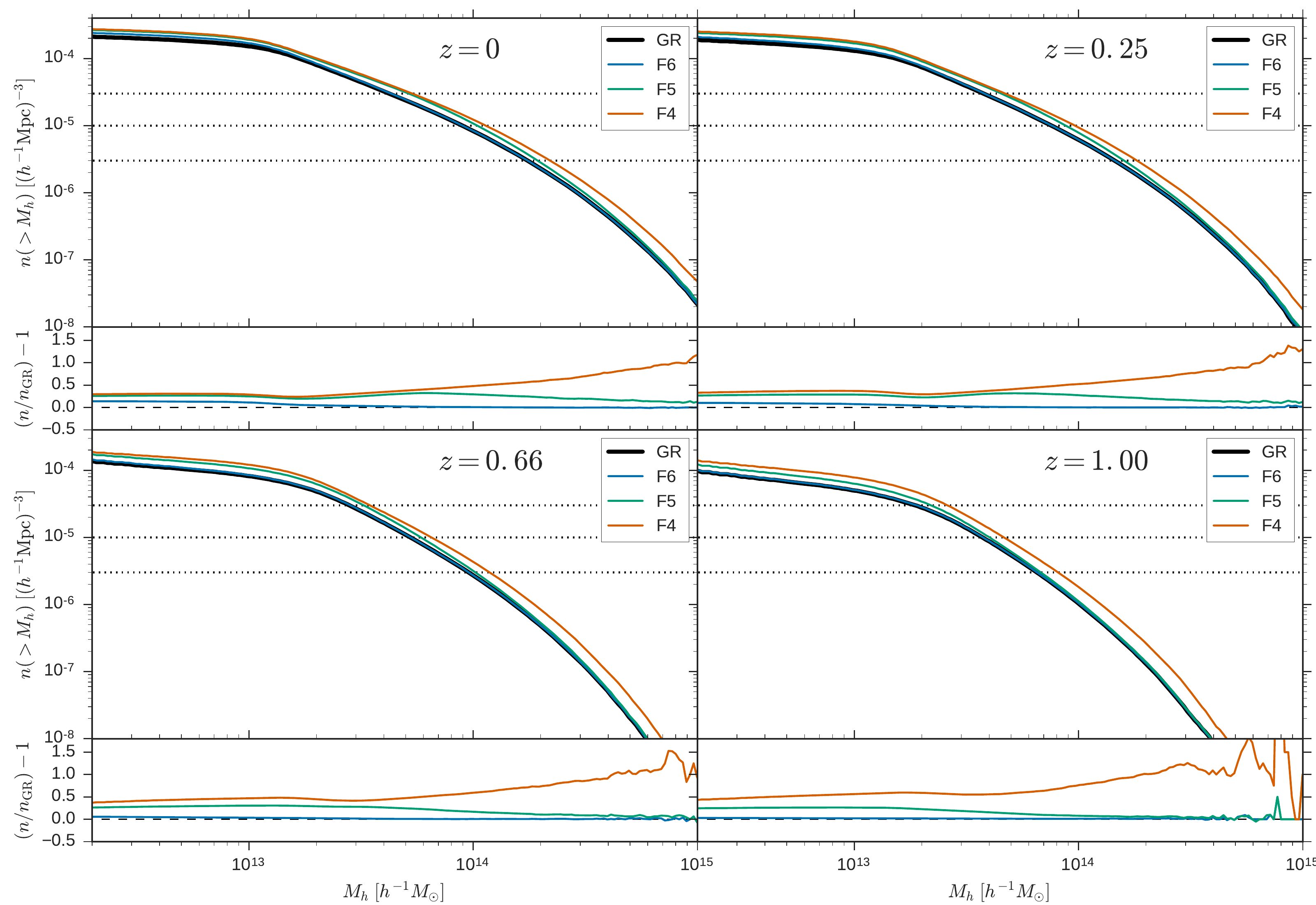}
  \caption{Cumulative halo mass function of the different models considered in this work for 
    the four different epochs $z=0, 0.25, 0.66, 1.0$, as indicated.
    The horizontal dotted lines signal the number densities we use to define our three halo samples.
    In each case, the lower panel show the relative change with respect to the $\lcdm$ (GR) model.}
  \label{fig:mass-func}
\end{figure*}

Before discussing the halo populations selected for our analysis, we need to consider the halo mass function of our simulations.
This is shown for four different times ($z=0,0.25,0.66,1.0$) in the four panels of Fig.~\ref{fig:mass-func}. 
It is quite obvious that our simulations suffer significantly from numerical shot-noise effects at the low mass end. 
Due to limited mass and spatial resolution of the simulations, the small-mass haloes suffer from the well 
known overmerging effect \citep{Klypin99,Klypin99b,Moore1999}. 
Thus the number density of small-mass haloes is underestimated. 
This is clearly indicated by the change of slope of the halo mass functions around $M_{200}\sim2\times10^{13}\Msun$. 
For GR, this mass roughly corresponds to  $\bar{n}=10^{-4} \mpcch$ at $z=0$, and 
to $\bar{n} = 3\times10^{-5} \mpcch$ at $z=1$. 
Although it seems that the magnitude of the resolution effects is very similar in all the models we 
study \citep{MOG_comp}, for the sake of fair comparison we decide to restrict ourselves to this limiting number density 
as the highest one we consider.
As the mass function is always larger for the $f(R)$ models than for GR, this limiting $\bar{n}$ should 
be sufficient for all our models.

The additional analysis of the plots in Fig.~\ref{fig:mass-func} reveals the behaviour already found by 
other authors \citep{sloh2009, Li2013, Hellwing2013}.
The largest deviation with respect to the $\lcdm$ case is observed, as expected, for the F4 model.
In this case, the mass function already shows a significant deviation from $\lcdm$ at $z=1$, with this deviation 
slightly increasing towards the largest halo masses.
The F5 model, with a more efficient screening, experiences a more complicated behaviour of the halo mass function. 
Due to the screening, the deviation is very small at the high-mass end, and we observe that the mass at which 
the halo abundances depart from the fiducial model is growing with time. 
For both the F4 and F5 models, for the range of halo masses not strongly affected by the screening, 
the relative departure of the halo number density from the $\lcdm$ case tends to shrink with time. 
This reflects the known effect that initially the $\lcdm$ model experiences a structure formation that is retarded
with respect to the MOG models, but at late evolutionary stages the halo growth slows down in the fifth-force 
cosmologies and so the $\lcdm$ is able to shrink the initial gap \citep{Hellwing2010}. 
This is mostly due to the relative scarcity of small haloes available for mergers, that is handicapping 
the halo mass growth via mergers at late times in $f(R)$.
Finally, for the F6 model we do not observe any significant deviation from the $\lcdm$ case.

We select different halo populations from our simulations by defining a series of \textit{threshold samples}, i.e. selecting haloes with mass above a certain value $M_{\rm min}$.
Since for massive haloes the virial mass - luminosity relation (or mass-to-light ratio) is monotonic 
and deterministic \citep{Moser2010}, such cuts are equivalent, on a first approximation, to a sample 
of galaxies selected by luminosity.
However, as the virial halo mass is not an observable, a selection with a fixed $M_{\rm min}$ can not 
be directly replicated in a real galaxy sample.
Instead, we decided to set a fixed number density $\bar{n}(\HP)$ for each of our samples, and define 
$M_{\rm min}$ in each model to match it. 
This approach is in essence a very simple version of the halo abundance matching.
As shown in Fig.~\ref{fig:mass-func}, the halo mass function can be significantly different in $f(R)$ 
models and in $\lcdm$.
This means that, for each sample defined in this way, we may end up with significantly different 
values of $M_{\rm min}$ in each of our models.

\begin{table}
  \caption{Properties of the halo samples used in this work. In each case, we list the minimum halo 
  mass $M_{\rm min}$ used to obtain the required number density $\bar{n}$ for a given redshift $z$. }
  \label{tab:samples}
  \centering
\resizebox{\columnwidth}{!}{
  \begin{tabular}{@{}ccc|rrrr@{}}
    \hline 
    & Halo & $\bar{n}$ & \multicolumn{4}{|c|}{$M_{\rm min} \, [10^{13} \Msun]$}  \\ 
    $z$ & population & $[\mpcch]$ & GR      & F6      & F5      & F4  \\ \hline \hline
    $0$ & $\HP_1$ & $3\times10^{-5}$       & $4.23$  & $4.31$  & $5.15$  & $5.36$  \\
        & $\HP_2$ & $10^{-5}$              & $8.94$  & $8.98$  & $10.43$ & $11.39$ \\ 
        & $\HP_3$ & $3\times10^{-6}$       & $17.50$ & $17.52$ & $19.41$ & $22.07$ \\ \hline
 $0.25$ & $\HP_1$ & $3\times10^{-5}$       & $3.77$  & $3.81$  & $4.54$  & $4.71$ \\
        & $\HP_2$ & $10^{-5}$              & $7.58$  & $7.62$  & $8.77$  & $9.63$ \\
        & $\HP_3$ & $3\times10^{-6}$       & $14.51$ & $14.51$ & $16.02$ & $18.26$ \\ \hline
 $0.66$ & $\HP_1$ & $3\times10^{-5}$       & $2.81$  & $2.83$  & $3.25$  & $3.48$ \\
        & $\HP_2$ & $10^{-5}$              & $5.23$  & $5.26$  & $5.84$  & $6.51$ \\
        & $\HP_3$ & $3\times10^{-6}$       & $9.57$  & $9.59$  & $10.26$ & $11.85$ \\ \hline
 $1.00$ & $\HP_1$ & $3\times10^{-5}$       & $1.86$  & $1.91$  & $2.20$  & $2.60$ \\
        & $\HP_2$ & $10^{-5}$              & $3.73$  & $3.75$  & $4.00$  & $4.65$ \\
        & $\HP_3$ & $3\times10^{-6}$       & $6.53$  & $6.55$  & $6.80$  & $8.10$ \\ \hline
  \end{tabular}
}
\end{table}

We defined three halo populations for the present work, $\HP_1, \HP_2, \HP_3$, with corresponding number 
densities $\bar{n} = 3\times10^{-5}$, $10^{-5}$ and $3\times10^{-6} \mpcch$, respectively.
The upper limit for the number density, $\bar{n}(\HP_1)$, was chosen based on the resolution limits of 
the simulations described above.
The lower limit $\bar{n}(\HP_3)$ was chosen to ensure that shot noise would not dominate our results.
These three number densities were used to select the corresponding samples at each of the redshift snapshots used.
Table~\ref{tab:samples} lists the corresponding values of $M_{\rm min}$ used in each case.
Following the differences in the mass function shown in Fig.~\ref{fig:mass-func}, the values are 
nearly identical for the $\lcdm$ and F6 models, while it is larger for F5 and F4.
As expected, in all cases, for a fixed $\bar{n}$ the corresponding $M_{\rm min}$ increases with decreasing redshift.

The number densities of the selected halo samples can be used to relate them to possible tracers to 
be used in the analysis of real surveys.
The density of $\HP_1$, for instance, is similar to that of the brightest samples of luminous 
red galaxies (LRGs) typically used in the analysis of the Sloan Digital Sky Survey \citep[SDSS, e.g.][]{Martinez2009b,Kazin2010b}.
Lower densities as those of $\HP_2, \HP_3$ are typical of galaxy groups or clusters of varying richness \citep{Koester2007a}.

\section{Results}
\label{sec:results}

In this section we present and discuss our main results obtained for the correlation function 
of dark matter and various halo samples at different epochs, both in position and in redshift space. 
We study the different components affecting the clustering of haloes separately. 
In section~\ref{ssec:matter-field} we study the clustering of the underlying matter density field, while 
in section~\ref{ssec:halo-corr-funct} we analyse the correlation function of haloes, derive 
the halo bias and assess its properties in the different models.

\subsection{Clustering properties of the matter density field}
\label{ssec:matter-field}

\begin{figure*}
  \centering
  \includegraphics[width=0.98\textwidth]{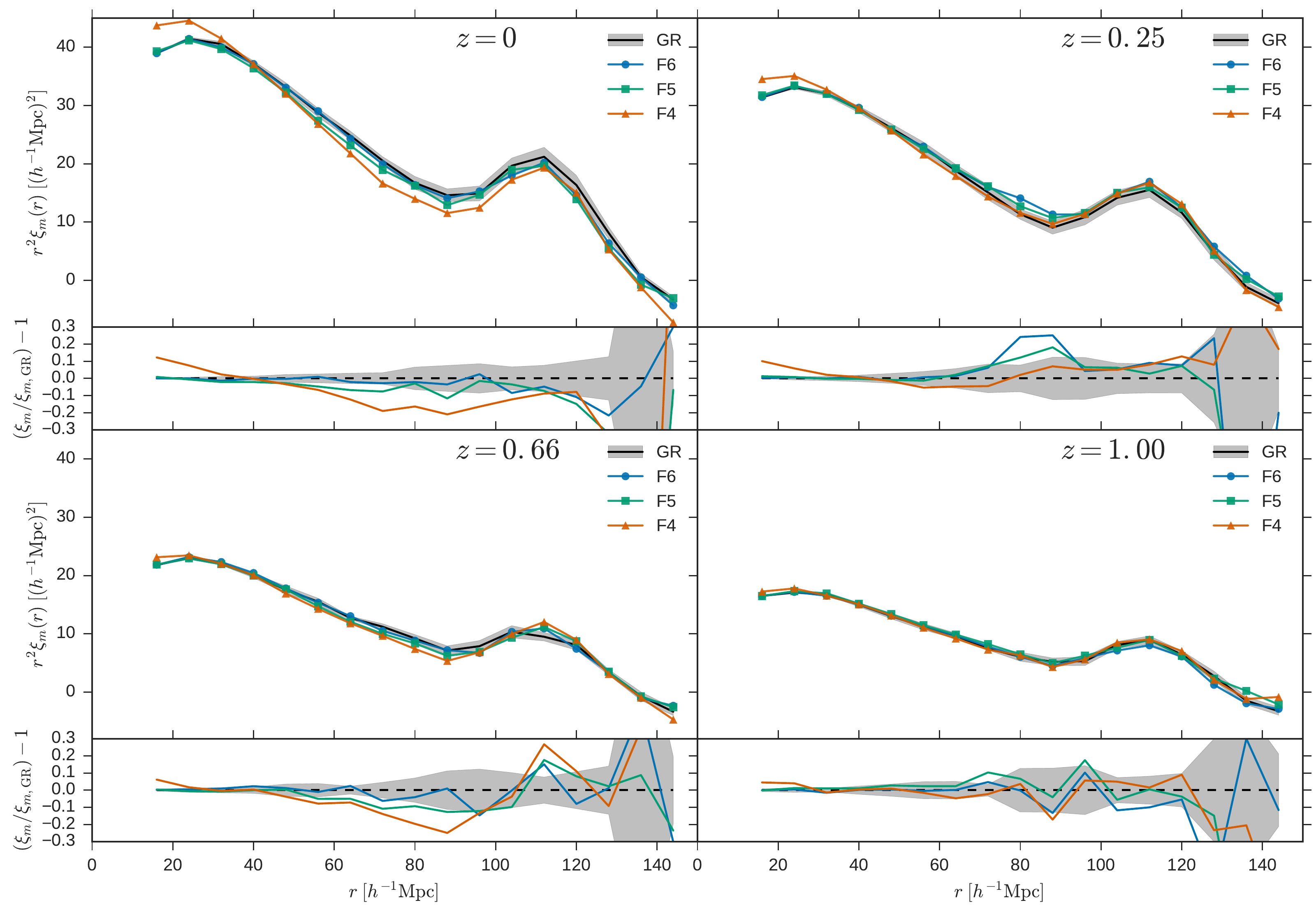}
  \caption{Real-space correlation function of the matter density field, $\xi_m(r)$, for the four models considered.
For the GR case we plot both the mean value (black line) and the $1\sigma$ scatter (shaded area) over the 6 realizations. For the different $f(R)$ models we only plot the corresponding mean values (colour lines and symbols, as indicated). The corresponding scatter is of the same order and scale-dependence as the GR one, hence we omit it for clarity. We plot the correlation function scaled by $r^2$ to better visualize the function at large scales, where its amplitude is low. Each plot corresponds to a different epoch, as indicated. In each case, the lower panels show the relative differences with respect to the GR case.
}
  \label{fig:xir_matter}
\end{figure*}

We first study the clustering of the smooth density field of the underlying matter component
of our simulations. Although this statistic is not directly accessible via astronomical observations,
it is noteworthy to study the properties of $\xi_{m}$, since it can be related and interpreted 
in a straightforward manner to the underlying theoretical model.
Figure~\ref{fig:xir_matter} presents the real-space correlation functions of the dark matter distributions
at the four redshifts considered. In each case, the black line and shaded area correspond 
to the mean value and $1\sigma$ scatter for the $\lcdm{}$ model. The corresponding scatter
of the modified gravity runs is of the same order and scale-dependence as the fiducial
$\lcdm{}$ case and hence it is not shown explicitly in the plot for clarity.
The different points and colour lines correspond to the three $f(R)$ models considered. 
The bottom panels in each case show the relative difference of the three $f(R)$ models with respect to the $\lcdm$ case.

We can already infer a number of interesting points from the data shown in Fig.~\ref{fig:xir_matter}. 
Firstly, we observe that the amplitude of clustering grows on all scales monotonically with cosmic time.
This is a well known result observed in all classes of cosmologies with hierarchical initial cold dark matter
power spectra.

One important feature illustrated by Fig.~\ref{fig:xir_matter} is the fact that the baryon acoustic 
oscillations (BAO) peak scale is not affected by modified gravity. 
It is apparent from the plot that all models show this peak at a scale $r_{\rm peak} \simeq 110 \hmpc$.
This corroborates our expectation that the expansion history of the $f(R)$ models is identical to that of $\lcdm$ when the ratio $c_1 / c_2$ is fixed according to eq.~(\ref{eq:c1c2ratio}).
However the position of the peak, and hence the cosmological information that can be extracted from it, could in principle be affected by non-linear effects acting differently in GR and $f(R)$ models.
To test this, we did a fit to our results using the simple model commonly used to analyse observations from galaxy redshift surveys \citep[see, e.g.,][]{Anderson2014a}. This model accounts for the non-linear damping of the BAO through the parameter $\Sigma_{NL}$, and measures a possible change in the BAO scale with respect to the fiducial value through the parameter $\alpha$.
We find that we recover the correct value of $\alpha = 1$ (and hence of the BAO scale) to within $2\%$ without any significant difference between models.
We do not find either any significant difference for $\Sigma_{NL}$, with values typically in the range $\Sigma_{NL} = 7 - 12 \hmpc$.

While the BAO peak scale is preserved, we can clearly notice in Fig.~\ref{fig:xir_matter} that all four considered models experience growth of clustering that differ from each other, with differences varying in magnitude and scales at which they appear.
At relatively early times the scalaron fifth force did not had enough time to significantly alter the growth 
of structures. 
This is clearly indicated by the results in the bottom-right panel, where at $z=1$ all models show matter 
clustering consistent with each other. 
However as the cosmic evolution progresses, we can observe a weak change of the correlation function amplitude in the $f(R)$ models. 

The F4 model at $z=0$ manifests a large excess at $r\la 35\hmpc$ when compared to $\lcdm$ and the two other $f(R)$ models. 
This is followed by a lower amplitude of $\xi_m$ in the regime from $r\ga 50\hmpc$ up to the BAO peak. 
This behaviour reflects the fact that the $f(R)$ models, and especially F4 (which is only very weakly screened), 
are characterized by a scale-dependent growth rate $f\equiv \dd \ln D_+/\dd \ln a$ \citep{Koyama2009}.
Such a strongly enhanced matter clustering at small scales comes with a price of matter that was more effectively 
evacuated from the interiors of large cosmic voids \citep{halosvoids_fr,Cai2015}. 
The overall effect is very strong in F4, which is indicated by a significantly altered slope of $\xi_m$ at $20\la r/(\hmpc)\la 90$.
The F5 model shows a much weaker discrepancy with respect to the GR case.
There is a hint of the amplitude of $\xi_m$ being lower than that of $\lcdm$ at scales of $r \sim 60 \hmpc$, similar to the F4 case.
At smaller scales, however, the F5 model results follow closely those of $\lcdm$, as expected from the stronger 
screening  in this case.
Similar behaviour (with weaker discrepancies) is found generally for the F4 and F5 models at $z=0.66$ and $z=0.25$.
At $z=0.25$ we see that the relative amplitudes of the MOG models versus $\lcdm$ at intermediate scales are 
slightly larger than expected from the global trends.
However this is a small variation that could be due to a statistical fluctuation.
The matter clustering of the F6 model is, at all redshifts, consistent with the $\lcdm$ case.

\begin{figure*}
  \centering
  \includegraphics[width=0.98\textwidth]{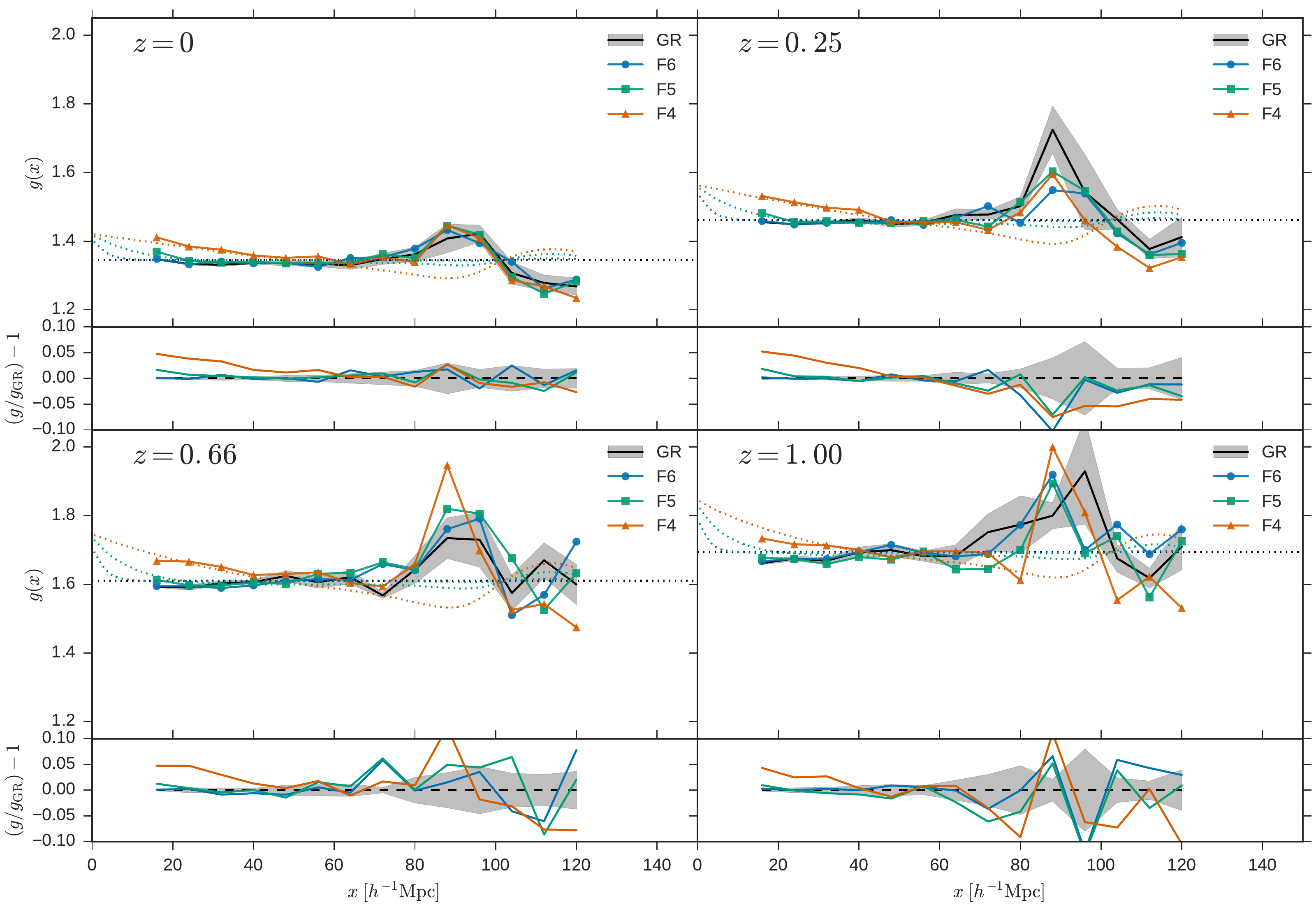}
  \caption{Ratio of the redshift-space to the real-space matter correlation functions, $g(x) = \frac{\xi_m^s(s=x)}{\xi_m^r(r=x)}$.
    As in previous figures, each plot corresponds to a different epoch, and lower panels show the relative differences with respect to GR.
    The shaded area corresponds to the $1\sigma$ scatter for the GR case.
    The dotted lines in the main panels illustrate the linear theory prediction $g_L(x)$ for each model. 
      The horizontal black line shows the constant prediction for $\lcdm$ according to eq.~(\ref{eq:Kaiser_growth_rate}), while the other
    lines show the scale-dependent predictions for F6 (blue), F5 (green) and F4 (orange) obtained using eqs.~(\ref{eq:1}, \ref{eq:3}).
    }
  \label{fig:zr-ratio_matter}
\end{figure*}

To study the effect of redshift-space distortions in the matter density field, we plot in 
Fig.~\ref{fig:zr-ratio_matter} the ratio of the redshift to real space correlation functions $g(x) \equiv \xi_m^s(s=x)/\xi_m^r(r=x)$. 
The results here can be compared to those of \citet{Jennings2012}, who studied the effect 
of redshift-space distortions in $f(R)$ cosmologies using power spectrum statistics for the same 
set of simulations as used in this work. 

In each case, we compare our results to the corresponding linear-theory predictions.
For the $\lcdm$ model, this corresponds to a constant ratio $g$, given by the Kaiser formula \citep{Kaiser1987,Hamilton1992},
\begin{equation}
  \label{eq:Kaiser_growth_rate}
  g_L^{GR} = 1 + \frac{2}{3} f + \frac{1}{5}f^2 \, ,
\end{equation}
where  $f$, the linear growth rate,\footnote{Not to be confused with the non-linear Lagrangian function $f(R)$}
can be approximated in $\lcdm$ by $f\approx \Omega_{\rm M}^{0.55}(z)$.
For our cosmogony $g\approx 1.35$ at $z=0$, growing to $g\approx 1.69$ at $z=1$. 
For the case of the $f(R)$ models, however, the growth rate depends on scale so the predicted ratio does also depend on scale.
\cite{Koyama2009} computed the corresponding Fourier-space growth rates for each of our models as function of wavenumber $f(k)$. 
We use these to compute the configuration-space linear prediction $g_L(x)$ for each model as follows. 
First, from the real-space linear power spectrum $P_L^r(k)$ we obtain the corresponding redshift-space power spectrum $P_L^s(k)$ using the Kaiser formula,
\begin{equation}
  \label{eq:1}
  P_L^s(k) = \left[1 + \frac{2}{3} f(k) + \frac{1}{5}f(k)^2 \right] P_L^r(k) \, .
\end{equation}
We use the $P_L^r(k)$ for each model and redshift obtained by \citet{Koyama2009}.
We obtain the corresponding real- and redshift-space linear correlation functions using the standard Fourier transform
\begin{equation}
  \label{eq:3}
  \xi_L^{r,s}(x) = 4 \pi \int_0^{+\infty} P_L^{r,s} (k) \frac{\sin(k x)}{k x} \frac{k^2 \dd k}{\left( 2 \pi \right)^3} \, ,
\end{equation}
and compute the linear prediction for the $g(x)$ ratio directly as $g_L(x) = \xi_L^s(s=x)/\xi_L^r(r=x)$.
In each panel of Fig.~\ref{fig:zr-ratio_matter} we show as dotted lines the linear-theory predictions calculated in this way for GR and our three MOG models.

Figure~\ref{fig:zr-ratio_matter} illustrates that on scales $ x \la 80 \hmpc$ the ratios $g(x)$ for 
all models follow remarkably well the corresponding linear predictions in each case.
This may seem to be in contradiction with the results of \citet{Jennings2012} which showed clearly the damping of the clustering due to virial motions at small scales ($k \gtrsim 0.05 \mpch$, see e.g. their fig.~4).
However, note that in this work we only consider scales $x \geq 10 \hmpc$.
In the case of $\lcdm$, we expect this damping effect to appear only at smaller scales in configuration space \citep[see, e.g.,][]{Scoccimarro2004}.
Our results indicate that this is the case also for the $f(R)$ models.

The results for F5 and F6 in Fig.~\ref{fig:zr-ratio_matter} agree to a good approximation with the measured $g(x)$ for $\lcdm$.
This is due to the fact that the excess clustering predicted by the scale-dependent growth rates of $f(R)$ appears typically at scales $x \la 20 \hmpc$ for these models -- as shown by the linear theory predictions (dotted lines)--, while we study mostly larger scales. 
In fact, we can observe that F5 deviates from GR at the smallest bin studied, $x = 16 \hmpc$.
F4 is clearly an outlier here, showing a clear enhancement in the ratio $g(x)$ with respect to GR at scales $x \la 50 \hmpc$.
This was to be expected: as we have already mentioned this model is nearly unscreened, hence its growth rate is larger than the $\lcdm$ one over a large range of scales. 
In the case of F5 and F6, on the other hand, the screening mechanism makes the deviations in the growth factor $f(k)$ with respect to GR to appear only at smaller scales ($k \gtrsim 0.01 \mpch$), as shown by fig.~1 of \citet{Jennings2012}.

The analysis of the redshift- to real-space matter correlation function ratios also reveals 
interesting behaviour around the BAO feature. 
In real space, the BAO feature has the form of a relatively sharp peak in the correlation function 
centred at the BAO scale (see Fig.~\ref{fig:xir_matter}).
In redshift space  the peculiar velocities introduce an smoothing of this BAO feature.
This means that the amplitude near the centre of the peak is reduced, and this power is moved to 
the scales corresponding to the tails of the peak.
When plotting the ratio $g(x)$ as in Fig.~\ref{fig:zr-ratio_matter} this results in the observed 
dip centred at the BAO scale ($x \simeq 110 \hmpc$), and a peak at slightly smaller scales ($x \simeq 90 \hmpc$).
This behaviour is observed for the four models --GR and $f(R)$-- considered.
The apparent larger differences between all models that appear at the peak and dip scales are artificially 
enhanced due to noise, since we take here a ratio of two very small quantities.
In this case, the noise is not expected to be Gaussian, so the simple error estimation we used 
does not fully account for it.

\subsection{Clustering of haloes}
\label{ssec:halo-corr-funct}

\begin{figure*}
  \centering
  \includegraphics[width=0.98\textwidth]{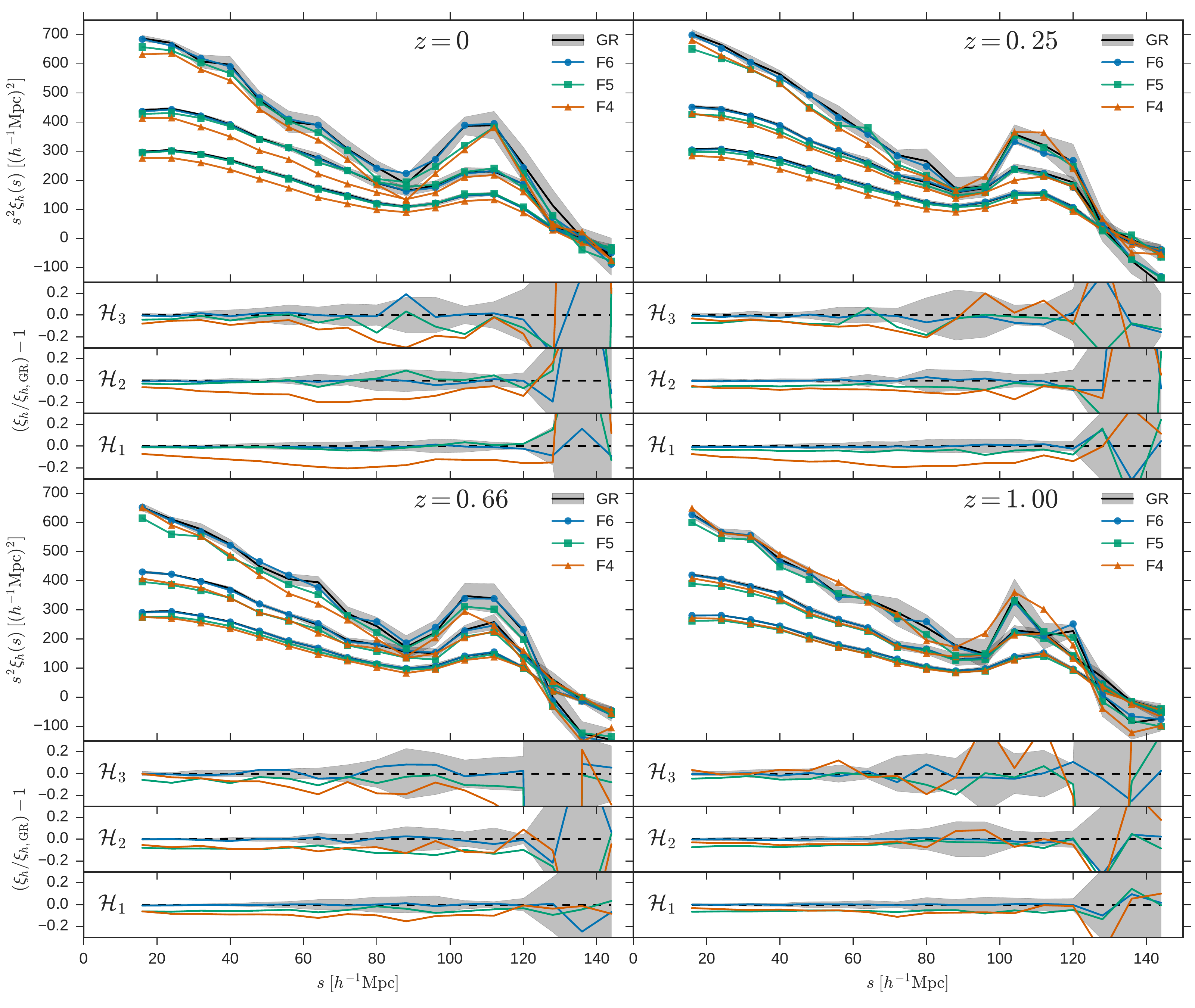}
  \caption{Redshift-space correlation functions of the different halo samples considered, $\xi_h(s)$, for our four models. 
As in Fig.~\ref{fig:xir_matter}, the functions amplitudes  were rescaled by $s^2$.
As in previous figures, each plot corresponds to a different redshift and the shaded area corresponds to the $1\sigma$ scatter for the GR case. 
In the main panels the different groups of lines correspond, from bottom to top, to 
the halo populations $\mathcal{H}_1, \mathcal{H}_2$ and $\mathcal{H}_3$.
The three lower panels in each plot show the relative differences with respect to GR for 
the indicated halo population. 
}
  \label{fig:cf-halos}
\end{figure*}

Now we turn to analyse the clustering properties of DM haloes.
We computed the redshift-space correlation functions $\xi_h(s)$ for our three halo populations 
$\HP_1, \HP_2$ and $\HP_3$ described in section~\ref{ssec:select-halo-popul}. 
We show our results for the four redshifts considered in Fig.~\ref{fig:cf-halos}.
Each of the main panels show the correlation function for the three populations in the four gravity models considered.
The lower panels show the relative difference for the three $f(R)$ models with respect to $\lcdm$, 
separately for each halo population.
As explained in section~\ref{sec:descr-halo-clust}, the $\xi_h(s)$ we computed correspond to the statistic 
that can be measured from samples of luminous galaxies or galaxy groups and clusters in real observations.
We could therefore compare directly our theoretical results with observational data.
Hence, any significant difference we see in Fig.~\ref{fig:cf-halos} can in principle serve as a way 
to discriminate between $\lcdm$ and the $f(R)$ models.

The four panels of Fig.~\ref{fig:cf-halos} (corresponding to four different redshifts) show the same 
main property of the clustering of our three halo populations in the four models considered. 
The least abundant halo samples ($\HP_3$) show the highest amplitude of the correlation function, 
while inversely the highest number density sample ($\HP_1$) exhibits the lowest $\xi_h(s)$ amplitudes, 
with the intermediate sample lying in between. 
This is expected, as higher number density corresponds to lower (average) halo mass 
(see Table~\ref{tab:samples}) and hence weaker clustering (lower bias parameter).

We focus now on the differences between models observed in the different $\xi_h(s)$.
For the four considered epochs, the F6 model is very close to the GR data. 
Hence the clustering of the three halo populations is statistically indistinguishable in these two models. 
F4 and F5 show more pronounced differences in all cases.
These differences can be appreciated more clearly (especially for F4) for populations $\HP_1$ and $\HP_2$.
This is partly due to the fact that for population $\HP_3$ the number of tracers is low hence 
the statistical error is the highest.
In general, we observe that the halo correlation functions for these two MOG models are significantly 
lower than the corresponding $\lcdm$ ones for scales $s \la 60 - 80 \hmpc$ 
(or even larger scales in some cases for F4).
It is interesting to note the difference in behaviour of the F5 and F4 models. 
For F5 the departure from the GR signal consists typically of a global change in the amplitude of 
$\xi_h(s)$ for each halo population and redshift.
This change in amplitude is visible for redshifts $z \geq 0.25$ but disappears at $z=0$.
In the case of F4 the deviation from GR seems to grow monotonically with time reaching the maximum at $z=0$. 
Moreover, in addition to the change in amplitude, we also observe for F4 a change in slope at scales 
$s \la 60 \hmpc$ with respect to the $\lcdm$ case.
This difference in slope is most clearly visible also at the lowest redshifts $z \leq 0.25$.
These differences in the clustering of haloes across redshifts reflect most likely a combination 
of many different effects: variations in the overall matter clustering (Fig.~\ref{fig:xir_matter}), 
discrepancies in the magnitude of the redshift space distortions (Fig.~\ref{fig:zr-ratio_matter}) and finally a deviation
in halo bias (see eq.~\ref{eq:linear_bias}).
To infer deeper into this, below we study the halo bias characteristics of our halo populations 
and the differences between our models.

\begin{figure*}
  \centering
  \includegraphics[width=0.98\textwidth]{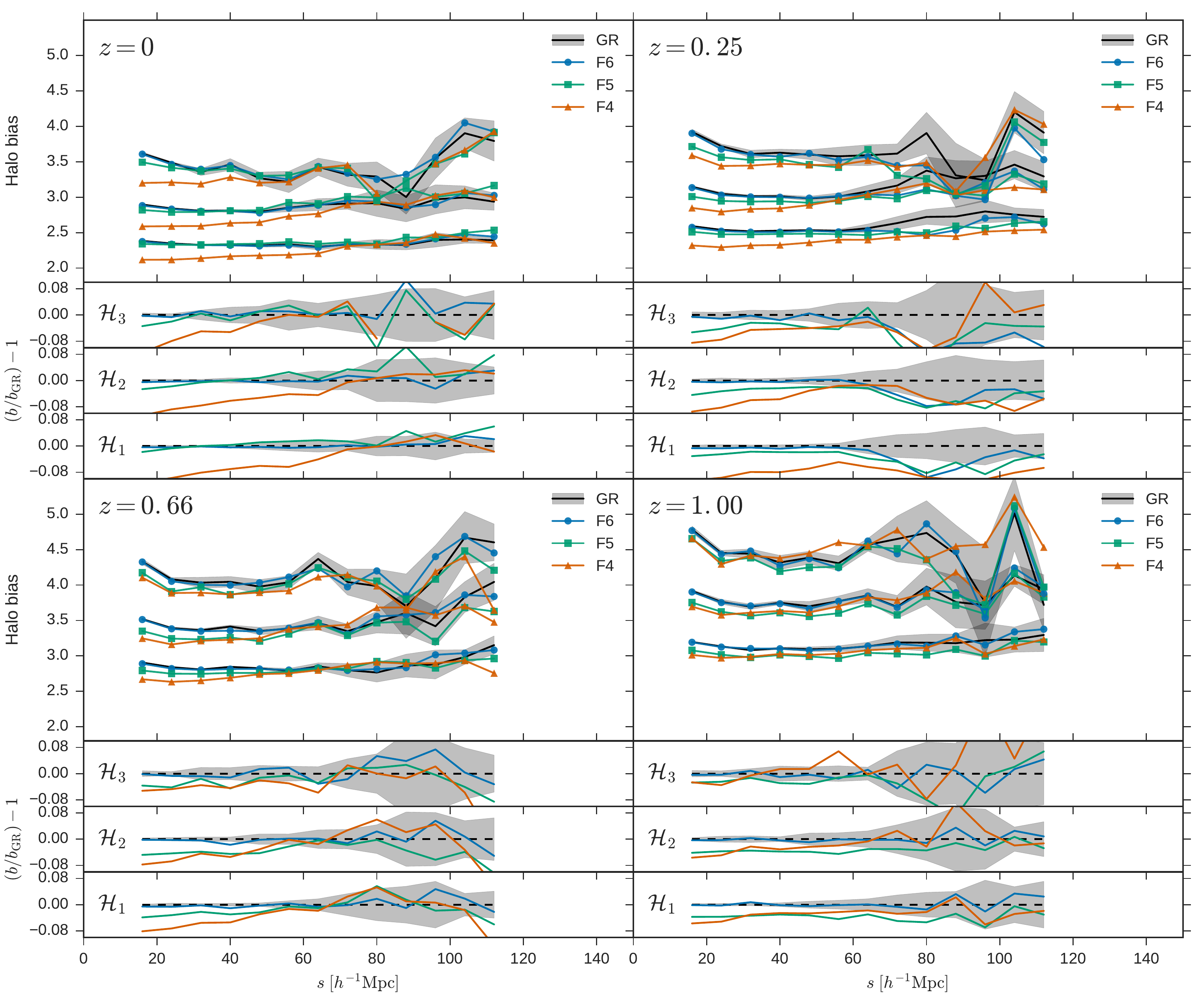}
  \caption{Redshift-space halo bias as function of scale $s$ for different halo populations. 
As in previous figures, each plot corresponds to a different redshift and the shaded area corresponds to the $1\sigma$ scatter for the GR case. 
In the main panels the different groups of lines correspond, from bottom to top, to 
the bias of halo populations $\mathcal{H}_1, \mathcal{H}_2$ and $\mathcal{H}_3$.
The three lower panels in each plot show the relative differences with respect to GR for 
the indicated halo population. }
  \label{fig:bias-scale}
\end{figure*}

\begin{figure}
  \centering
  \includegraphics[width=\columnwidth]{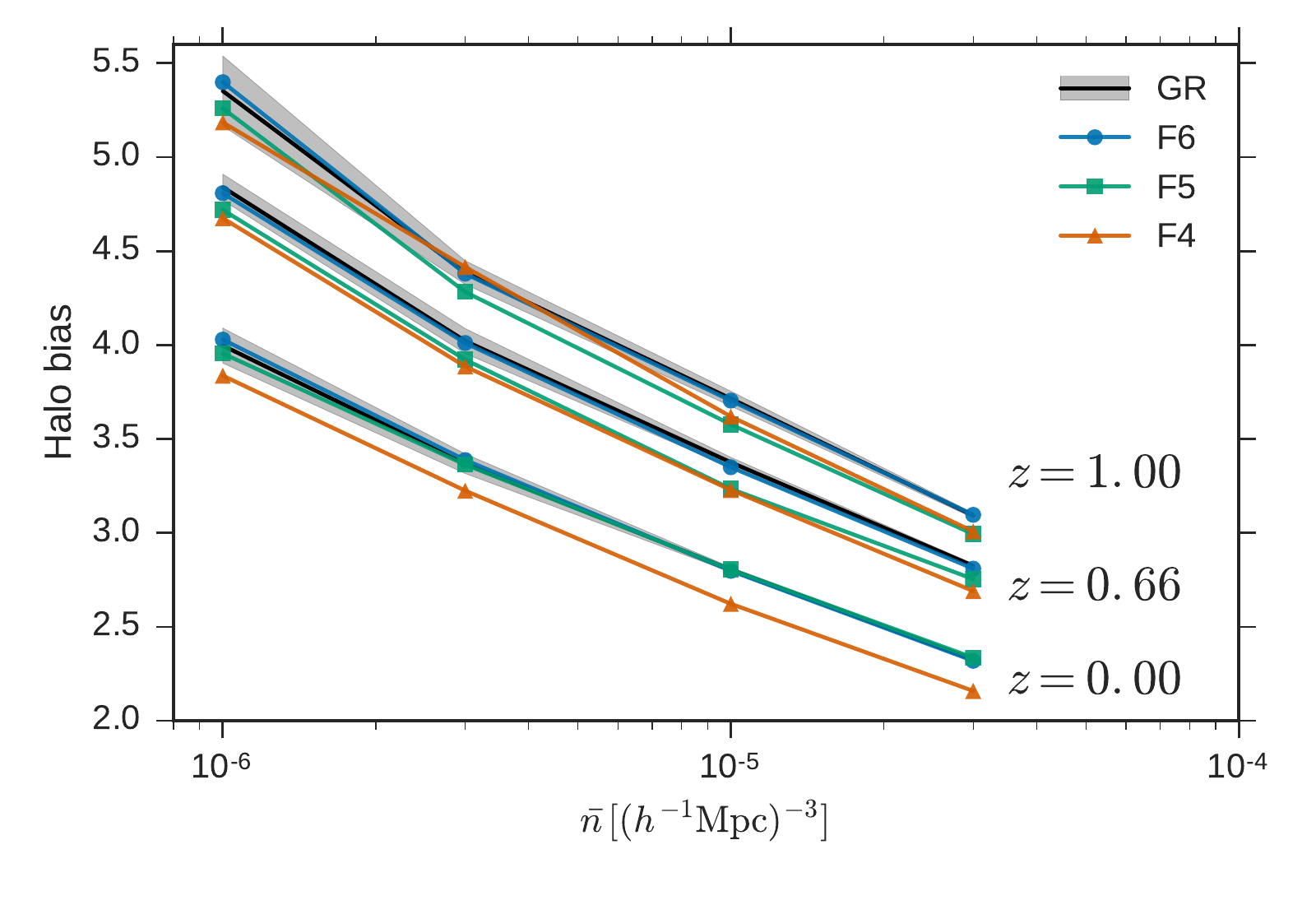}
  \caption{Redshift-space halo bias estimated over the range $s \in [24,52] \hmpc$ 
   as function of the halo density used for selection of the populations. 
   In addition to the samples used elsewhere in this work, we also show for completeness 
   the result for the sample with $\bar{n} = 10^{-6} \mpcch$.
   The results for three different redshift snapshots, $z=0, 0.66$ and $1.0$, are shown 
   as indicated by the labels.
   We omit here the results for $z=0.25$ for clarity.}
  \label{fig:bias-haloes}
\end{figure}

Following eq.~(\ref{eq:linear_bias}), we measure the redshift-space halo bias from the simulations using the estimator
\begin{equation}
  \label{eq:4}
  b_h(s) = \sqrt{\frac{\xi_h(s)}{\xi_m(s)}} \, .
\end{equation}
We focus here only on the redshift-space bias as this is the theoretical quantity relevant 
for comparison with typical observational measurements of $\xi$ in galaxy surveys. 
However our results for the real-space bias are very similar to the ones presented here, 
with only a global change in the amplitude.

Figure~\ref{fig:bias-scale} presents the bias as a function of scale $s$ for our three halo populations. 
We only plot $b(s)$ for $s \leq 110 \hmpc$ to avoid scales where either $\xi_m(s)$ or $\xi_h(s)$ become negative.
As a complementary plot we also show in Fig.~\ref{fig:bias-haloes} the bias,  averaged over a range 
of scales $s \in [24,52] \hmpc$, as a function of the number density of the halo population. 
We chose these scales as in that range the bias is reliably measured and approximately constant 
in the case of $\lcdm$, as shown in Fig.~\ref{fig:bias-scale}.
At smaller scales the non-linear evolution of the density and velocity fields becomes important 
and a simple linear bias description breaks down. 
The rapidly increasing $b(s)$ value observed at $s \la 20\hmpc$ is a hint of this non-linear behaviour. 
The weak scale dependence observed at large scales, where $s>80\hmpc$, may be also due to 
non-linear effects near the BAO peak.
However, at these scales the scatter is large due to cosmic variance and low amplitudes of both 
matter and halo $\xi(s)$, so these effects are not statistically significant here.

The $f(R)$ halo samples are characterized always by a bias that is either smaller or equal to the fiducial GR case. 
If we recall that in $f(R)$ haloes tend to be, on average, more massive than in GR, this result may seem surprising. 
If we look at a population of haloes at fixed virial mass, in $f(R)$ there will be many haloes 
that originate from smaller density peaks than their equivalent $z=0$ mass cousins in GR. 
Since the initial conditions are the same within the ensemble, haloes that originate from 
smaller peaks (lower Jeans mass), which are characterized by lower bias, are in $f(R)$ shifted 
towards higher masses and then compared with the fiducial GR case that originate from rarer peaks (hence higher bias).
This is consistent with the picture seen in Fig.~\ref{fig:bias-haloes}, where we observe that differences in bias
are higher for higher redshift, just as the differences in mass functions in Fig.~\ref{fig:mass-func}.
Another feature of the $f(R)$ halo bias seen in Fig.~\ref{fig:bias-scale} is its stronger scale
dependence than in the $\lcdm$ case. This can be especially seen for F4, where at $20\leq s/(\hmpc)\leq 60$
the bias is increasing with scale. Similar but much weaker behaviour can be also observed
for F5 at $z=0.66$.

Once we have studied the differences in halo bias, we can better interpret the differences in the halo correlation function $\xi_h(s)$  observed between the $f(R)$ models F4 and F5 and the $\lcdm$ case (Fig.~\ref{fig:cf-halos}).
For the case of the F4 model, it is clear that the steeper slope and excess clustering at small 
scales observed for both $\xi_m(r)$ (Fig.~\ref{fig:xir_matter}) and $g(x)$ (Fig.~\ref{fig:zr-ratio_matter}) 
is compensated by a significantly smaller bias (and positive scale dependence).
This results in the $\xi_h(s)$ having in all cases a smaller amplitude than the $\lcdm$ case, 
but with only a mild change of slope (except at $z=0$).
A similar explanation can be given for the halo correlation functions in the F5 model, although 
the differences with respect to $\lcdm$ in this case are smaller.

\section{Observational tests using clustering statistics}
\label{sec:observ-tests}

The results shown in section~\ref{ssec:halo-corr-funct} indicate that for both F4 and F5 halo samples of 
the same number density $\bar{n}$ can be characterized typically by significantly different values 
of the correlation function $\xi_h(s)$ than our fiducial GR model. 
In principle, as our halo samples can be directly related to samples of luminous galaxies or groups 
in real surveys (with the caveats discussed in section~\ref{sec:descr-halo-clust}), these $\xi_h(s)$ are observable quantities.
Hence, they could be used to discriminate between GR and these MOG models.
However, the differences in $\xi_h(s)$ seen in Fig.~\ref{fig:cf-halos} could be degenerate with changes 
in the $\lcdm$ clustering due to variations of the cosmological parameters, and in particular $\sigma_8$.
Therefore, one would need to combine the $\xi_h(s)$ measurements with other model-independent 
determinations of these parameters.
An alternative would be to measure directly the halo bias and use the differences between models seen 
in Figs.~\ref{fig:bias-scale} and \ref{fig:bias-haloes}.
Bias can not be directly obtained from two-point statistics, but there exist estimates based on 
weak lensing observations \citep{McKay2001,Covone2014,vanUitert2016} and on higher-order statistics 
of the galaxy distribution \citep[e.g.][]{Verde2002a,Gaztanaga2005a, McBride2011a,Arnalte-Mur2016a}.
However, these methods infer bias from observations in a model-dependent way, hence all 
the systematic effects were checked only against the assumed $\lcdm$ cosmology. 

In this section we explore a way in which we can nevertheless use the two-point clustering of haloes 
to test observationally the studied $f(R)$ models. 
We try to define a statistic based on the clustering of haloes that: (i) can differentiate between 
$\lcdm$ and different $f(R)$ models, following our results in section~\ref{ssec:halo-corr-funct}, 
and (ii) can be measured from observations in a way which is as model independent as possible 
(i.e. does not depend on the clustering properties of the matter density field).
As the differences between models observed above vary with both scale and halo population, 
the best way to define such a statistic is to combine the correlation functions $\xi_h(s)$ 
for different populations and at different scales.

Hence, we define  the relative clustering ratio $\CR$ for a halo population $\HP$ as function of scale $s$ as
\begin{equation}
  \label{eq:2}
  \CR(s, \HP| \HP_{\rm ref},  s_{\rm ref}) = \frac{s^2 \xi_h(s|\HP)}{s_{\rm ref}^2 \xi_h(s_{\rm ref}|\HP_{\rm ref})} \, ,
\end{equation}
where $\HP_{\rm ref}$ is a reference halo population and $s_{\rm ref}$ is a reference scale (kept fixed). 
Here we use the term $s^2/s_{\rm ref}^2$ to rescale the correlation functions in order to have
comparable values as a function of $s_{\rm ref}$.
As we show below, this new statistic can be predicted theoretically for each model using the results of section~\ref{ssec:halo-corr-funct}.
It can also be directly measured from observations -- with the caveats mentioned in section~\ref{sec:descr-halo-clust} to identify a halo population with a class of observed objects.
The way to compute $\CR$ in a given survey is to first identify the relevant populations equivalent to $\HP$ and $\HP_{\rm ref}$ and obtain the corresponding catalogues of objects. 
One then computes the redshift-space correlation function for each of these catalogues using an standard estimator \citep[e.g.][]{Landy1993a}.
The clustering ratio $\CR$ is finally computed using directly eq.~(\ref{eq:2}) above.
In this way, $\CR$ will be independent of the amplitude of the matter correlation function, $\sigma_8$.
Furthermore, as both populations $\HP$ and $\HP_{\rm ref}$ are extracted from the same volume (same survey), the effects of sampling variance of the $\CR$ ratio will be additionally suppressed.

Here, we choose for the reference population $\HP_{\rm ref} = \HP_1$, the sample with 
the highest spatial abundance, $\bar{n} = 3\times10^{-5} \mpcch$.
The scale-dependent differences between models in $\xi_h(s)$ can appear in different ways in $\CR$ depending on the value of $s_{\rm ref}$ used.
Therefore $s_{\rm ref}$ can be chosen, in principle, to maximize the differences between models.
Here, we show our results for two reference scales: $s_{\rm ref} = 16$ and $64  \hmpc$.
These two values were chosen to span the range of scales where the discrepancies are more clearly observed in Fig.~\ref{fig:cf-halos}, while avoiding larger scales where errors can grow significantly.

\begin{figure*}
  \centering
  \includegraphics[width=0.98\textwidth]{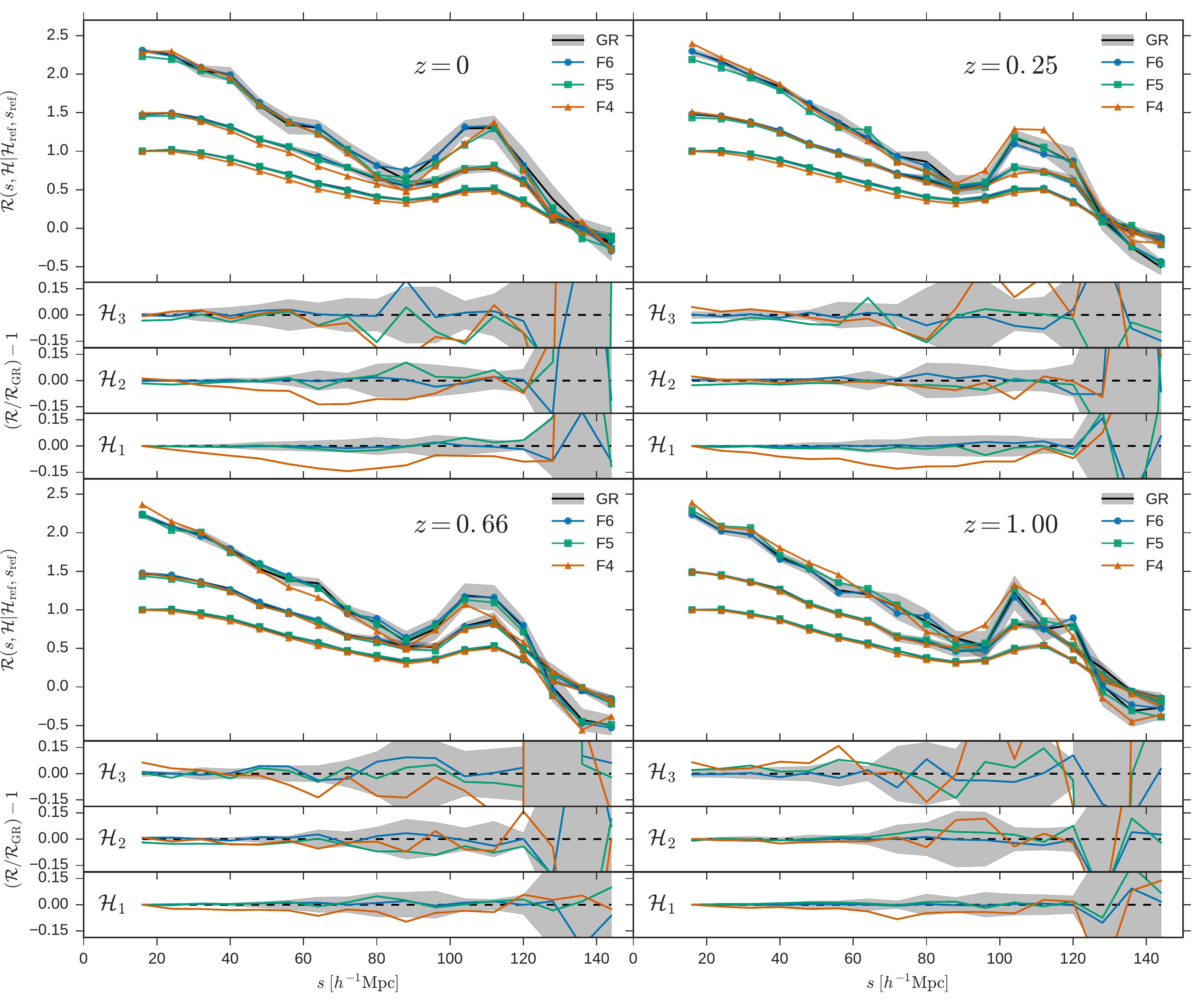}
  \caption{Relative clustering ratio $\mathcal{R}(s,\HP|\HP_{\rm ref}, s_{\rm ref})$ for different halo populations, for the case in which the reference sample is $\HP_{\rm ref} = \HP_1$,   and the reference scale is set to $s_{\rm ref} = 16 \hmpc$. 
As in previous figures, each plot corresponds to a different redshift and the shaded area corresponds to the $1\sigma$ scatter for the GR case. 
In the main panels the different groups of lines correspond, from bottom to top, to 
the clustering ratio obtained for halo populations $\mathcal{H}_1, \mathcal{H}_2$ and $\mathcal{H}_3$.
The three lower panels in each plot show the relative differences with respect to GR for 
the indicated halo population.
}
  \label{fig:clustratio_s16}
\end{figure*}

\begin{figure*}
  \centering
  \includegraphics[width=0.98\textwidth]{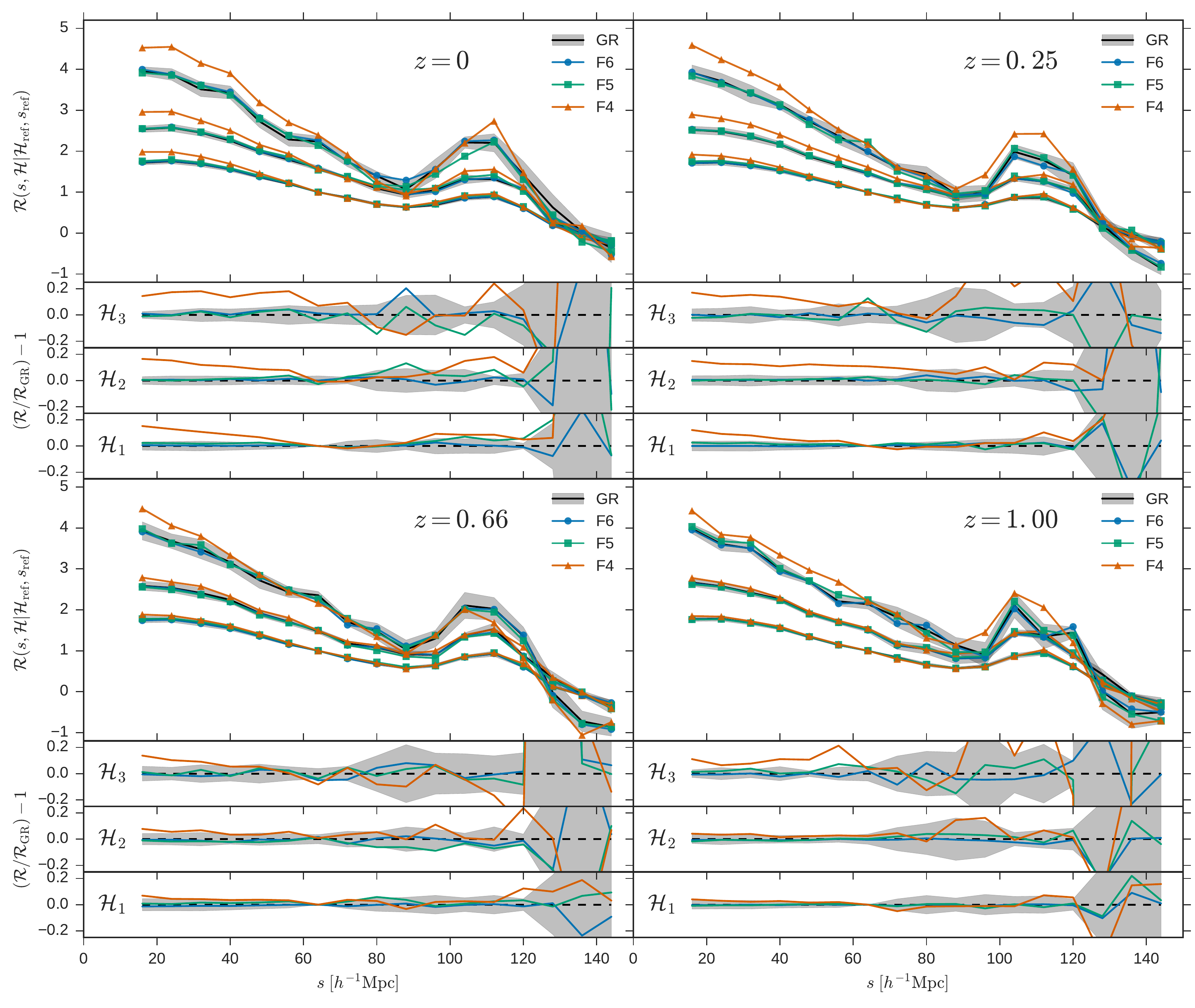}
  \caption{Same as Fig.~\ref{fig:clustratio_s16}, for the case in which the reference scale is set to $s_{\rm ref} = 64 \hmpc$.}
  \label{fig:clustratio_s64}
\end{figure*}

In Figs.~\ref{fig:clustratio_s16} and \ref{fig:clustratio_s64} we plot the clustering ratio $\CR$ 
for our two reference scales, $s_{\rm ref}= 16\hmpc$ and $s_{\rm ref}=64\hmpc$ respectively.
The clustering ratio $\CR$ for the $\HP_1$ population is a special case, as this is the population 
we use as reference for our calculations.
In this case $\CR(s)$ is just the halo correlation function $\xi_h(s)$ normalized to its amplitude 
at $s=s_{\rm ref}$. 
As the difference between the halo correlation functions of F5 and GR was just a constant shift 
in the amplitude, this difference completely disappears in the case of $\CR$. 
For the F4 model, however, this difference with respect to GR has a dependence on scale, and 
therefore we also see a significant deviation in $\CR$ at $z=0, 0.25$ for both values of $s_{\rm ref}$.

When we consider different samples ($\HP_2$ and $\HP_3$) the situation changes, as here the $\CR$ 
depends also on the relative bias between different populations.
For $s_{\rm ref} = 16 \hmpc$ (Fig.~\ref{fig:clustratio_s16}), F5 presents some departures from GR, 
exhibiting lower values of $\CR$ for $s \la 40\hmpc$ at $z \leq 0.66$.
These departures are small (only a few per cent), but significant for $\HP_2$.
As these scales (and $s_{\rm ref}$) correspond to the mildly non-linear regime, this could be due 
to the non-linear effects scaling differently with halo mass in F5 and GR.
This discrepancy could be used, in principle, to discriminate between the F5 model and $\lcdm$. 
However, given the small size of the effect, this would be difficult in practice due, e.g., to possible systematic errors.

Moving to the larger reference scale $s_{\rm ref} = 64 \hmpc$ (Fig.~\ref{fig:clustratio_s64}) 
the results for F5 are completely consistent with GR. 
This indicates that the F5 signature at linear scales is reduced to a global change in 
the amplitude of clustering.
On the other side, we obtain here deviations from GR that are large and statistically significant 
for the least screened $f(R)$ model, F4. 
These deviations grow with decreasing redshift, attaining relative changes of $\sim 20 \%$ at 
the smallest scales for all halo populations.
For $z \leq 0.25$ the statistical significance of these deviations is $\sim 2-5\sigma$.
This indicates that $\CR(s, \HP | \HP_1, s_{\rm ref}=64\hmpc)$ can be used to render constraints 
for strongly deviating models like F4.
As expected from all our previous results, for all the considered snapshots and reference scales, 
the $\CR$ of F6 are statistically consistent with GR.

\section{Discussion and conclusions}
\label{sec:conc}

We have analysed the real- and redshift-space two-point clustering statistics
of DM and haloes in a series of simulations employing the structure formation 
in the $\lcdm$ and $f(R)$ cosmological models. We have also introduced a new statistic
 - the halo relative clustering ratios $\CR(s, \HP| \HP_{\rm ref},  s_{\rm ref})$.
We have fixed our analysis on three halo populations constructed by implementing
fixed number density cuts at $\bar{n}= 3\times10^{-5}$, $10^{-5}$ and $3\times 10^{-6}\mpcch$ 
(denoted, respectively, $\HP_1, \HP_2$ and $\HP_3$).
Hence our halo populations mimic in a general sense spatial selection effects similar to those found 
in volume-limited samples from redshift galaxy surveys. 
The number densities we use are typical of samples of very luminous galaxies, or of groups and clusters of galaxies.
We can summarize our most important findings in the following points:
\begin{itemize}
 \item In all models the clustering amplitude of DM grows monotonically with time.
       At high redshifts the matter clustering is indistinguishable among models.
       At later times ($z \la 0.66$) our strongest model - F4 - shows significant
       deviations of the amplitude and slope of $\xi_m$ at small and intermediate scales, while $\xi_m$
       of both F5 and F6 remain mostly consistent with $\lcdm$. In all models the BAO peak scale 
       is the same and is not affected in any significant way by the fifth force;
 \item The ratio, $g(x)$, of redshift- to position-space matter correlation functions
       of F5 and F6 is compatible at large-scales ($x\geq25\hmpc$) with the
       $\lcdm$ results. F4 is a strong outlier here, showing significant
       deviations up to $x\sim 50\hmpc$.
       All four models show good agreement with the respective linear theory predictions
       in the range $15 \la x / (\hmpc) \la 80$.
 \item The differences of the redshift-space two-point correlations of haloes are bigger
       than in the case of the DM density field. 
       In general, the halo correlation functions of the F4 and F5 models are lower than those
       of GR.
       This is more clearly observed for the $\HP_1$ and $\HP_2$ samples, because of the larger
       errors in $\HP_3$ (due to sparse sampling).
       The strong F4 model is an outlier at all epochs, with the strongest signal at $z=0$. 
       However, for the F5 model the $\xi_h$ reaches its maximal departure from GR 
       at intermediate and high redshifts, $z \geq 0.25$.
 \item Halo bias in all $f(R)$ models and for all halo populations is always 
       lower than in GR or consistent with the fiducial model. Again F4 is an outlier here
       at all scales and epochs. The F6 model halo bias is fully consistent with the GR
       predictions, while for F5 the most significant differences appear again at intermediate and high redshifts.
 \item Finally we considered the relative clustering ratios $\CR$ to construct a largely
       model-independent observational clustering probe. 
       The F4 model halo clustering ratios depart significantly from the GR model for all our samples, 
       specially when using as reference scale $s_{\rm ref} = 64 \hmpc$ and at $z  \leq 0.25$.
       Again F6 is characterized by too small differences from GR to be statistically distinguishable in any way.
       However the $\CR$ of the mild F5 model at redshifts of $z \leq 0.66$
       and for  $s_{\rm ref} = 16\hmpc$ is showing a small but significant signal at scales $s\la 40\hmpc$.
\end{itemize}

Our results indicate that only in the case of the unrealistically strong and not screened F4 model one 
can expect a clear, strong and significant signal visible in both the matter and halo clustering. 
This signal for F4 is also clear in the clustering ratios $\CR$.
This means that this model could be tested using only the two-point clustering of haloes in a model-independent way.
On the other end of the spectrum the highly screened F6 model is always very close to GR for all our statistics and samples and at all epochs.
Hence, these models are indistinguishable from each other, at least when one is concerned with 
the two point clustering statistics. 
For the physically interesting F5 model we have found only small differences with respect to GR 
in the clustering ratios $\CR$.
It shows, however, a significant signal in the raw halo correlation functions $\xi_h(s)$, that 
can be summarized as changes in a constant linear bias as function of halo population 
and redshift (Fig.~\ref{fig:bias-haloes}).
The predicted signal is strongest for redshifts $z \geq 0.25$.
Two-point clustering observations can not be used to measure the bias on their own.
However, our results suggest that they could be used in combination with other probes 
(e.g. an independent measurement of $\sigma_8$) to put constraints on the F5 model.

Our results yield the hope that growing observational data will be able to constrain this 
class of $f(R)$ models using galaxy clustering.
Two near-term projects that may have the potential to perform these tests are the DESI \citep{Levi2013,DESICollaboration2016a}
and J-PAS \citep{Benitez2014a} surveys, which will cover a large fraction of the sky ($14000 \deg^2$ and $8500 \deg^2$, 
respectively).
Both projects will target different classes of galaxies up to redshifts $z \la 1$, therefore covering 
the range of redshifts studied in this work. 
Given the expected number density, it will be possible to select samples of galaxies (e.g. LRGs) that can be related to the halo populations we studied.
It will also be possible to use for these tests catalogues of galaxy groups and clusters from these surveys.
\citet{Ascaso2016a} showed that it will be possible to detect reliably in J-PAS clusters corresponding to halo masses of $M\gtrsim 3.6\times10^{13} \Msun$ up to $z \simeq 0.7$ (assuming a $\lcdm$ cosmology).
This selection would match our $\HP_1$ sample at the lowest redshifts.
Slightly further in the future, another survey suitable for this type of analysis will be \textit{Euclid} \citep{Laureijs2011}. \textit{Euclid} will observe galaxies over a significantly larger volume, thus reducing the statistical error of the clustering measurements. However, its spectroscopic survey will be limited to higher redshifts than those studied in this work ($z \gtrsim 0.9$), where we expect the differences in clustering between the GR and $f(R)$ to be smaller. 

We therefore expect that our method will be able to constrain the particular \citet{hs2007} model considered here down to $\left| f_{R0}\right| \simeq 10^{-5}$ using data from these near-future surveys.
This is competitive with possible constraints using other known methods. 
\citet{DESICollaboration2016a}, for example, forecast that in the ideal case DESI will be able to measure the growth rate $f$ at scales $k \leq 0.1 \mpch$ to a precision of $\simeq 2 - 4 \%$ for redshifts $z \in [0.6, 1.0]$ (see their tables 2.3 and 2.4). 
For comparison, \citet{Jennings2012} show that the maximum expected change in $f$ with respect to GR at these scales and redshifts is $\simeq 5\%$ ($\simeq 1 \%$) for $\left| f_{R0}\right| \sim 10^{-5}$ ($\left| f_{R0}\right| \sim 10^{-6}$).
Therefore, this type of measurements could yield constraints of the same order as those achievable using the clustering ratios $\CR$.
Alternatively, \citet{Cataneo2015} predict that it will be possible to obtain even better constraints when future surveys allow for the detailed measurement of the cluster and group mass function to higher redshifts ($z \sim 2$).

Our analysis of the $ g(x) \equiv \xi_m^s(x)/\xi_m^r(x)$ ratios showed, however, that one needs to take 
caution when trying to extract the growth rate $f$ from just the halo/galaxy/matter clustering signal. 
Although the scale-dependent growth rates predict an enhancement in the redshift-space clustering in $f(R)$ models, this is only seen at relatively small scales, $x<20\hmpc$ ($x<50\hmpc$ in F4).
At these small scales, we will find deviations from linear theory due to the effect of virial motions, and one should take into account the predicted enhanced peculiar velocities in MOG models \citep{Zu2014,Hellwing2014PhRvL,Sabiu2016}.
That means that, in order to use the growth rate inferred from galaxy redshift catalogues to constrain this class of models, it is necessary to model these effects in detail, and to test the analysis method with realistic MOG mocks \citep{Barreira2016}.
This problem is largely alleviated in the case where we only consider redshift-space 
related quantities, such as the halo correlation function $\xi_h(s)$ or specially 
the clustering ratios $\CR$, and so avoid the necessity of modelling precisely 
the connection between position and redshift space objects.  
In sum we advertise here to use $\CR(s, \HP| \HP_{\rm ref},  s_{\rm ref})$ to 
study and constrain $f(R)$ models using the redshift space clustering of galaxies 
as measured by modern galaxy surveys.

\section*{Acknowledgements}
We thank the referee, Dr Nelson Lima, for his comments that helped improve the clarity of the paper.
The authors are very grateful to Baojiu Li for many inspiring comments and for providing cosmological simulations used in this work. 
We have also benefited from comments and discussions we had with Shaun Cole, Carlos Frenk, Kazuya Koyama, Will Percival and Violeta Gonz\'alez-P\'erez. 
WAH is grateful to the co-authors for the endless patience they showed during writing of this draft.
PAM was supported by the European Research Council Starting Grant (DEGAS-259586) and by the Generalitat Valenciana project PrometeoII 2014/060, and acknowledges additional support from the Spanish Ministry for Economy and Competitiveness through grants AYA2013-48623-C2-2 and AYA2016-81065-C2-2-P.
WAH acknowledges support from the European Research Council grant through 646702 (CosTesGrav) and the Polish National Science Center under contract \#UMO-2012/07/D/ST9/02785.
PN acknowledges the support of the Royal Society through the award of a University Research Fellowship, the European Research Council, through receipt of a Starting Grant (DEGAS-259586) and the Science and Technology Facilities Council (ST/L00075X/1).
This work used the DiRAC Data Centric  system at Durham University,
operated by the Institute for  Computational Cosmology on behalf of 
the STFC DiRAC HPC Facility (\url{www.dirac.ac.uk}). This equipment 
was funded by BIS National E-infrastructure capital grant ST/K00042X/1, 
STFC capital grant ST/H008519/1, and STFC DiRAC Operations grant 
ST/K003267/1 and Durham  University. DiRAC is part of the National 
E-Infrastructure. This research was carried out with the support of the 
HPC Infrastructure for Grand Challenges of Science and Engineering 
Project, co-financed by the European Regional Development Fund under 
the Innovative Economy Operational Programme.
This research made use of \textsc{matplotlib}, a Python library for publication quality graphics \citep{Hunter2007a}.

\bibliographystyle{mnras}
\bibliography{clustering_haloes_bib}
\bsp

\label{lastpage}
\end{document}